\documentclass[runningheads]{llncs}
\usepackage{amsmath,amssymb,amsfonts,amstext}
\usepackage{algorithm2e}
\usepackage{csquotes}
\usepackage{graphicx}
\usepackage{hyperref}
    \hypersetup{
        pdftitle={Implementing and Executing Static Analysis using LLVM and CodeChecker},    
        pdfauthor={G\'{a}bor Horv\'{a}th, R\'{e}ka Nikolett Kov\'{a}cs, Rich\'{a}rd Szalay, and Zolt\'{a}n Porkol\'{a}b},     
        pdfnewwindow=true,      
        hidelinks,              
        colorlinks=false,       
        linkcolor=red,          
        citecolor=green,        
        filecolor=magenta,      
        urlcolor=cyan           
    }
\usepackage{marvosym}
\usepackage[cachedir={.},frozencache]{minted}
    \setminted{autogobble}
    \setminted{bgcolor=backgroundgrey}
    \setminted{escapeinside=@@}
    \setminted{mathescape}
    \setmintedinline{bgcolor={}}

    \makeatletter
        \@ifpackagelater{minted}{2022/12/12}{
          \ifdefined\minted@optlistcl@quote
          \ifwindows
            \renewcommand{\minted@optlistcl@quote}[2]{%
              \ifstrempty{#2}{\detokenize{#1}}{\detokenize{#1="#2"}}}
          \else
            \renewcommand{\minted@optlistcl@quote}[2]{%
              \ifstrempty{#2}{\detokenize{#1}}{\detokenize{#1='#2'}}}
          \fi
          \fi

          \newcommand{\minted@def@optcl@novalue}[2]{%
            \define@booleankey{minted@opt@g}{#1}%
              {\minted@addto@optlistcl{\minted@optlistcl@g}{#2}{}%
               \@namedef{minted@opt@g:#1}{true}}
              {\@namedef{minted@opt@g:#1}{false}}
            \define@booleankey{minted@opt@g@i}{#1}%
              {\minted@addto@optlistcl{\minted@optlistcl@g@i}{#2}{}%
               \@namedef{minted@opt@g@i:#1}{true}}
              {\@namedef{minted@opt@g@i:#1}{false}}
            \define@booleankey{minted@opt@lang}{#1}%
              {\minted@addto@optlistcl@lang{minted@optlistcl@lang\minted@lang}{#2}{}%
               \@namedef{minted@opt@lang\minted@lang:#1}{true}}
              {\@namedef{minted@opt@lang\minted@lang:#1}{false}}
            \define@booleankey{minted@opt@lang@i}{#1}%
              {\minted@addto@optlistcl@lang{minted@optlistcl@lang\minted@lang @i}{#2}{}%
               \@namedef{minted@opt@lang\minted@lang @i:#1}{true}}
              {\@namedef{minted@opt@lang\minted@lang @i:#1}{false}}
            \define@booleankey{minted@opt@cmd}{#1}%
                {\minted@addto@optlistcl{\minted@optlistcl@cmd}{#2}{}%
                  \@namedef{minted@opt@cmd:#1}{true}}
                {\@namedef{minted@opt@cmd:#1}{false}}
          }

          \minted@def@optcl@novalue{customlexer}{-x}
        }{
          \minted@def@optcl@novalue{customlexer}{}
        }
    \makeatother
\usepackage{pifont}
    \newcommand{\tick}{\ding{51}}
    \newcommand{\cross}{\ding{55}}
\usepackage{rotating}
\usepackage{subcaption}
    \DeclareCaptionSubType{listing} 
\usepackage{tikz}
    \usetikzlibrary{arrows}
    \usetikzlibrary{positioning}
    \usetikzlibrary{shapes}
    \tikzstyle{box}=[draw, minimum size=1cm, text width=1.8cm, text centered]
\usepackage[obeyFinal,shadow]{todonotes}
    \setuptodonotes{color=yellow!50, fancyline, textcolor=blue, tickmarkheight=.5cm}
\usepackage[capitalise]{cleveref} 
    \crefname{sublisting}{listing}{listings}
    \Crefname{sublisting}{Listing}{Listings}
\usepackage{xcolor}
    \definecolor{americanrose}{rgb}{1.0, 0.01, 0.24}
    \definecolor{armygreen}{rgb}{0.29, 0.33, 0.13}
    \definecolor{backgroundgrey}{rgb}{0.95, 0.95, 0.95}
    \definecolor{babyblueeyes}{rgb}{0.63, 0.79, 0.95}
    \definecolor{darkgray}{rgb}{0.66, 0.66, 0.66}
    \definecolor{darkspringgreen}{rgb}{0.09, 0.45, 0.27}
    \definecolor{green(ryb)}{rgb}{0.4, 0.69, 0.2}
    \definecolor{lavenderpurple}{rgb}{0.59, 0.48, 0.71}
    \definecolor{royalpurple}{rgb}{0.47, 0.32, 0.66}
    \definecolor{spirodiscoball}{rgb}{0.06, 0.75, 0.99}
    \definecolor{white}{rgb}{1.0, 1.0, 1.0}

\newcommand{\CC}{C\nolinebreak\hspace{-.05em}\raisebox{.4ex}{\tiny\bf +}\nolinebreak\hspace{-.10em}\raisebox{.4ex}{\tiny\bf +}}

\begin{document}
\title{Implementing and Executing Static Analysis using LLVM and CodeChecker\thanks{%
    This work presented in this paper was supported by the European Union, co-financed by the European Social Fund in project \emph{EFOP-3.6.3-VEKOP-16-2017-00002}.}%
}
\titlerunning{Implementing and Executing Static Analysis using LLVM and CodeChecker}
%
\author{G\'{a}bor Horv\'{a}th\orcidID{0000-0002-0834-0996} \and
    R\'{e}ka Nikolett Kov\'{a}cs\orcidID{0000-0001-6275-8552} \and
    Rich\'{a}rd Szalay\orcidID{0000-0001-5684-5158} \and
    Zolt\'{a}n Porkol\'{a}b\orcidID{0000-0001-6819-0224}}
\authorrunning{G.\ Horv\'{a}th et al.}
%
\institute{Department of Programming Languages and Compilers, \\
    Institute of Computer Science, Faculty of Informatics, \\
    E\"{o}tv\"{o}s Lor\'{a}nd University, Budapest, Hungary \\
    \email{\{hogtabi,kovacs.reka,szalayrichard,gsd\}@inf.elte.hu}
}
\maketitle              
\begin{abstract}
Static analysis is a method to analyse the source code without executing it.
It is widely used to find bugs and code smells in industrial software.
Among other methods, the most important techniques are the one based on the abstract syntax tree and the one performing symbolic execution.
Both of these methods found their role in modern software development as they have different advantages and limitations.
In this tutorial, we present two problems from the \CC{} programming language: the elimination of redundant pointers; and how can we deal with dangling pointers originated from the misuse of the \mintinline{CPP}{std@$::$@string} class.
These two issues have different theoretical backgrounds and finding them requires different implementation techniques.
We will provide a step-by-step guide to implement the checkers -- software to identify the aforementioned problems --, one is based on the abstract syntax analysis method, the other will explore the possibilities of the symbolic execution.
The methods are explained in great detail and supported by code examples.
We intend this tutorial both for architects of static analysis tools and for those developers who want to understand the advantages and the constraints of the individual methods.

\end{abstract}

\section{Introduction}

Static analysis is the analysis of a program based solely on the source text, without execution, carried out by an automated tool.
It is widely used for problems such as code optimisation, large-scale refactoring, finding code metrics, and code visualisation.
In spite of the strong theoretical limits on its power (e.g.\ the halting problem, as explained in \cref{handout:challenges}), static analysis proved to be a valuable means of code quality assurance and continues to be an active field of research.

Software maintenance costs increase with the size of the code-base.
Fortunately, static analysis was found to play a great role in reducing the maintenance costs of complex software products~\cite{Zhivich2009}.
For example, compilers can detect more and more optimisation possibilities statically.
These optimisations enable developers to use high-level language features without degrading the performance of the program, making software development less error-prone and generally easier.
Static analysis is also a great approach for finding bugs and code smells~\cite{Bessey2010}, and can act as a cheap supplement to testing.
This is important because bugs found in production are significantly more costly to fix~\cite{fixcost}.
Additionally, some properties of the code - such as compliance with coding conventions - cannot be tested, but can be enforced by static analysis.

CodeChecker~\cite{CodeChecker} is an open source project to integrate different static analysis tools such as the Clang Static Analyzer and Clang-Tidy into the build system, CI loop, and the development workflow.
It also has a powerful issue management system to make it easier to evaluate the reports of the analysis tools.

The open-source \emph{LLVM/Clang} compiler~\cite{lattner2008llvm} has two components that are doing user-facing static analysis: \emph{Clang-Tidy}~\cite{clangtidy}, and the \emph{Clang Static Analyzer}~\cite{static-analyzer}.

Clang-Tidy is an open source tool employed extensively within Google to find code smells and refactor code.
It uses a domain-specific pattern matching language on the abstract syntax tree.

The Clang Static Analyzer implements a powerful symbolic execution engine for C, \CC{}, and Objective-C, together with several useful checks to detect programming errors.
\emph{Symbolic execution}~\cite{Hampapuram2005} is a major \emph{abstract interpretation} technique.
During symbolic execution the program is being interpreted without any knowledge about the run-time environment.
Symbols are used to represent fix but unknown values (the value might be only known at run-time, e.g.\ user input), and calculations are carried out on them symbolically.
Throughout the process, the analyser attempts to enumerate all possible execution paths.
Due to the vast number of possible paths, it uses heuristics to decide where to cut the analysis, how to deal with missing information, and how to approximate certain aspects of the program.

This paper is intended to give a step by step introduction for the above methods.
We will explain the technical background for both approaches, examine real problems to solve and provide thorough code examples for the implementation details.
The rest of the paper is organized as follows:
In \cref{overview} we overview the technical foundations of the most popular static analysis methods.
In \cref{codechecker} we give a brief introduction of CodeChecker, an open source static analysis tooling system which runs various bug finding methods.
In the rest of the paper we describe the two main methods to write bug finding checkers in details.
The Syntax Tree-based analysis for the Clang-Tidy is discussed in \cref{tidy}.
The symbolic execution based method is explained in \cref{sa}.
The paper concludes in \cref{conclusion}.

\section{An Overview of Static Analysis}\label{overview}

\subsection{Compilers}

Nowadays, compilers are used for various use-cases.
We find multiple compilers in browsers, graphics card drivers, your spreadsheet application, your machine learning framework, and so on.
Knowing the basics of how compilers work might help you understand the tools that you use on a daily basis.

Historically, compilers are the most popular users of static analysis.
Many of the algorithms and data structures that are used in static analysis tools originate from compiler research.
This section introduces some of the compiler related concepts that are used throughout this paper.

\paragraph{Lexical analysis}\label{handout:lexical-analysis}

The compilation process usually starts with \emph{lexical analysis} (\emph{lexing}).
During lexing, the compiler identifies lexical elements in the source text such as operators, identifiers, and literals.
The result of the lexing phase is a stream of \emph{tokens}.
Tokens are lexical elements that consist of a source location and a kind.

\paragraph{Parsing}\label{handout:parsing}

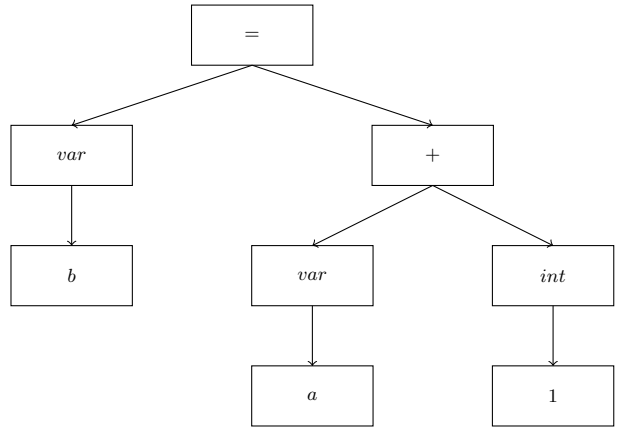
\begin{figure*}
    \centering
    \vspace{10pt}
    \begin{tikzpicture}[scale=0.8, transform shape]

        \node[box] (assign) at (-2,3) {$=$};

        \draw[->] (-2,2.5) -- (-5,1.5);
        \draw[->] (-2,2.5) -- (1,1.5);

        \node[box] (var) at (-5,1) {$var$};
        \node[box] (plus) at (1,1) {$+$};

        \draw[->] (-5,0.5) -- (-5,-0.5);
        \draw[->] (1,0.5) -- (-1,-0.5);
        \draw[->] (1,0.5) -- (3,-0.5);

        \node[box] (b) at (-5,-1) {$b$};
        \node[box] (var2) at (-1,-1) {$var$};
        \node[box] (int) at (3,-1) {$int$};

        \draw[->] (-1,-1.5) -- (-1,-2.5);
        \draw[->] (3,-1.5) -- (3,-2.5);

        \node[box] (a) at (-1,-3) {$a$};
        \node[box] (one) at (3,-3) {$1$};
    \end{tikzpicture}
    \vspace{10pt}
    \caption{The abstract syntax tree of the expression $b = a + 1$.}\label{handout:ast}
\end{figure*}

The token stream is then \emph{parsed}.
During parsing, the compiler builds an \emph{abstract syntax tree} (\emph{AST}).
This tree data structure represents relationships between lexical elements.
The syntax tree is called abstract because some of the lexical elements might not appear explicitly (like parentheses), because this information may be available implicitly in the structure of the tree.
It may also contain implicit constructs that do not appear in the source text.
On the other hand, \emph{parse trees} or \emph{concrete syntax trees} contain all the information about the source text (including white space and comments), so the original text can be restored from this data structure using pretty-printing.
It is common not to build a parse tree at all, only build the abstract syntax tree instead.
A parse tree can be useful when precise pretty printing support is required.
An example AST can be viewed in \cref{handout:ast}.

After parsing the source code, certain syntactic constructs are sometimes reduced to a simplified form.
This method is called \emph{desugaring}.
Basically, the language is reduced to a core language which is much simpler to deal with.
This core language has the same expressive power as the original one, but it is not as convenient to write.

A process closely related to desugaring is called \emph{cannibalisation}.
In some cases, there are more ways to express the same concept in a language or an intermediate representation, e.g.\ there might be many ways to represent loops.
Cannibalisation will try to transform different representations to the same form so later phases dealing with these constructs are less complex.

\paragraph{Semantic analysis}\label{handout:semantic-analysis}

After parsing, the next phase is \emph{semantic analysis}.
This phase consists of \emph{type checking} and \emph{scope checking}.
The AST is decorated with type information during this step.
The resulting data structure contains enough information to generate executable code.

In some languages, in order to resolve ambiguities in the grammar, it is necessary to have some type information available.
In this case, parsing and semantic analysis is done in the same phase.

\paragraph{Error handling}\label{handout:compiler-errors}

During compilation, errors might be found in the source text.
There are multiple kinds of possible errors depending on which phase is responsible to for finding them.
The most common ones are lexical errors, syntax errors, and type errors.

Nowadays compilers provide the user with advanced warning messages, e.g.\ they can detect variables that are uninitialised on an execution path.
In order to carry out such an analysis, it is often beneficial to build the \emph{control flow graph} (\emph{CFG}).
The CFG is a graph whose nodes are basic blocks, and the edges are the possible jumps between those blocks.
A \emph{basic block} is a code fragment that contains only sequentially executed instructions.
This means that it contains no jump instructions and each instruction can only jump to the beginning of a basic block.

Since creating a parser for a language requires heroic effort, it is common to build static analysis tools on the top of an existing compiler.
This way the tool uses the internal representation of the compiler, e.g.\ the AST or the CFG instead of introducing its own parser.

\subsubsection{LLVM}

The LLVM Compiler Infrastructure Project is a collection of modular and reusable compiler and tool-chain technologies.
Although the name \emph{LLVM} was originally an acronym of \emph{a Low-Level Virtual Machine}, the project has since outgrew the meaning of its long name, and now uses the short version exclusively.

The LLVM core libraries provide a modern source and target-independent optimiser and code generators for many popular architectures.

\begin{figure}
    \centering
    \includegraphics[width=\textwidth]{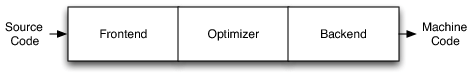}
    \caption{Steps taken by a simple compiler.}\label{handout:SimpleComp}
\end{figure}

Traditionally compilers have three parts.
The front-end is parsing the source code and creates an intermediate representation.
The middle end or optimiser does platform independent optimisations.
The back-end will generate the executable code.
This architecture can be seen in \cref{handout:SimpleComp}.

LLVM has its own intermediate representation.
There are multiple compilers that can generate LLVM IR.
These compilers are utilising the same optimisation and code generation libraries.
This approach significantly lowers the barriers writing a new industrial strength optimising compiler and also helps the community to focus on better tools by sharing the work on the middle end and the back-end among the developers of different compilers.

\begin{figure}
    \centering
    \includegraphics[width=\textwidth]{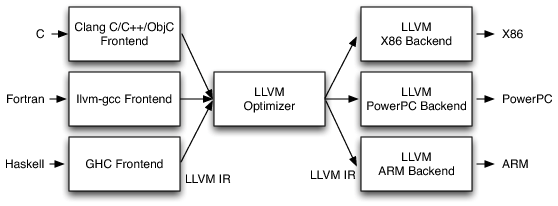}
    \caption{The architecture of the LLVM Compiler Infrastructure project.}\label{handout:LLVMComp}
\end{figure}

This architecture also improves the interoperability between languages.
Combining the LLVM IR from multiple front-ends might enable optimisations that span across calls between different languages.
The architecture can be seen on \cref{handout:LLVMComp}.

The optimiser is doing fair amount of static analysis on the LLVM IR.
Two of the major tasks of optimisations are the following: establishing a cost model and approximate the execution time of a piece of code and proving that a transformation does not change the semantics of the code.
If a transformation is safe and it improves the performance of the program according to the cost model the optimiser has, LLVM will apply the transformation.

There are lots of different transformations within LLVM and they are called \emph{passes}.
The order of passes has a significant effect on the end result.
It is also common to run a pass multiple times or have a pass whose sole purpose is to make code more suitable for the subsequent passes.

\subsection{Static analysis challenges}\label{handout:challenges}
Due to the limitations of static analysis, in general, it is inevitable to either have false positives (incorrectly reporting an error) or false negatives (missing errors)~\cite{chess2004static}.
In software verification and mission critical applications, it is more important to catch all possible errors.
For a practical bug finding tool, however, it is more important to have few false alarms, so developers will take the results seriously~\cite{johnson2013don}.
Usually, it is possible to reduce the number of false negative results at the cost of increasing the number of false positive ones and vice-versa.
This is captured by the so-called binary confusion matrix, which can be seen in \cref{handout:bug-classification}.

\begin{table}
    \centering
    \begin{tabular}{|l|c|r|}
        \hline
                          & Bug exists     & Correct code     \\ \hline
        Error reported    & True Positive  & False Positive   \\ \hline
        No error reported & False Negative & True Negative    \\
        \hline
    \end{tabular}
    \medskip
    \caption{Classification of the results of static analysis.}\label{handout:bug-classification}
\end{table}

The halting problem is the problem of determining, from a description of an arbitrary program and an input, whether the program terminates or runs forever. Since Alan Turing proved that no general program exists that solves the halting problem for all program-input pairs~\cite{turing1936computable}, we know that it is infeasible to automatically prove the correctness of large-scale programs. Therefore, analyses need to resort to approximations, which can result in false positive and false negative reports. There are several techniques to deal with false positives. It is possible to rank the results and make important violations appear first, or to use statistical methods and other heuristics.

Fortunately, the approximations employed by analyses work well most of the time. As programs containing complex data and control dependencies are also difficult for humans to understand, they are regarded as bad practice and generally avoided.

Rice's theorem from 1953~\cite{Rice:53} can be informally paraphrased as ``all interesting questions about program behaviour are undecidable''.
This can be easily proved for any special case~\cite{moller2012static}.
Assume the existence of an analyser that can check whether a variable has a constant value during the execution of the program.
We could use this analyser to decide on the halting problem by using an input program such as \cref{handout:const_check} where \mintinline{CPP}{tm} returns \mintinline{CPP}{true} if and only if the $j$th program in the enumeration terminates on empty input.

\begin{listing}
    \begin{minted}{CPP}
        int x = 5;
        if (tm(j))
            x = 6;
    \end{minted}
    \caption{Input for const-ness check}\label{handout:const_check}
\end{listing}

This can be a very discouraging result, but most of the time the real focus of static analysis is not to prove such properties but rather to solve practical problems.
This means that such analysis can often produce approximate results.

\subsubsection{Challenges implied by languages}

The syntax and the type system of a language can have a significant effect on the compilation time and the complexity of language-related tools~\cite{grossman2005cyclone}.
The compilation model has similar impacts.
Programs written in dynamically typed languages tend to convey much less information statically about the properties of the program, making the static analysis much more challenging~\cite{madsen2015static}.
In order to overcome this obstacle, it is common to augment static analysis with information that was gathered during an execution of the program.

The efficiency of static analysis tools depends on how easy it is to parse a language.
The main challenge in parsing is ambiguity~\cite{aho1975deterministic}.
In order to solve ambiguities, the parser needs to look ahead and read additional tokens.
The minimum number of times a token needs to be inspected in the worst case depends on the grammar of the language.
Designing the language grammar with these problems in mind can make the life of compiler and tool authors easier.

The grammar of C and \CC{} is ambiguous without sufficient type information.
To get the type information the headers included in a file need to be parsed before the file can be parsed.
We detail the issues with token-based analysis and the highlights of syntax-tree based analysis in \cref{tidy:ast}.

One of the factors that makes static analysis harder is the number of dynamic bindings~\cite{bacon1996fast}.
In case a virtual method is invoked, the analyser cannot know which method will be executed unless the allocation of the receiver object is known.

The type system also has an important role in the analysis. The purpose of the type system is to prove certain properties of the program - in this sense, the process of type checking is also static analysis.
Static analyses often assume that the analysed program is correctly typed, because it provides the analysis with a lot of useful information - the more expressive the type system, the more information is likely to be available at compile time.

Note that type checking has the same limitations as any other kind of static analysis.
Although there are strongly normalising type systems that guarantee the termination of all well-typed programs~\cite{barendregt1992lambda}, either the expressiveness of those systems is very limited or the type checking of an arbitrary program is undecidable.

\subsection{Static analysis methods}\label{handout:methods}

There are several methods to analyse software statically.
The easiest way is to transform the code into a canonical form, tokenise it, and then use regular expressions or other pattern matching method on that token stream.
This method is efficient and it is used in CppCheck~\cite{CppCheck}.
Unfortunately, it has a lot of weaknesses as well, for example, the type information cannot be utilised easily.
It is also hard to match for implicit constructs in the language, for example implicit casts.

A more advanced approach is to use the AST of the source code and match patterns on the tree representation.
The tree can be handled more easily than a token stream.
In case this AST is typed, type information and implicit constructs are available as well.
One of the  disadvantages of this approach is that it is hard to do any control flow-dependent matching on the AST.
For example, it is hard to reason about the values of the variables in this representation.
For some class of analysis this approach is sufficient.
Clang-Tidy is using this approach.
We detail syntax-tree based analysis in \cref{tidy}.

In a flow-sensitive analysis the analyser builds the CFG.
A flow-sensitive analysis is a polynomial analysis that traverses the CFG and analyses each node during the traversal.
Each time two branches fold into one, the information gathered on those branches are merged.
This analysis is powerful enough to catch issues like non-trivial cases of division by zero.

\begin{listing}
    \begin{minted}{CPP}
        int i;

        if (a)
          i = 0;
        else
          i = 1;

        // $\vdots$

        if (!a)
          j = 5 / i;
        else
          j = 3 / i;

        j += 2 / i;
    \end{minted}
    \caption{Example for division by zero.}\label{handout:div_zero}
\end{listing}

Unfortunately, flow-sensitive analysis is not precise enough for some purposes.
Consider an analysis that checks for division by zero errors.
It would track the values of the variables and store which ones are zero.
What should such analysis do, when two branches fold together?
One option would be to mark the value of a variable zero when it was zero on any of the branches.
If you look at \cref{handout:div_zero}, there will be a false positive case when $5$ is divided by \mintinline{CPP}{i}, since this branch is only taken when $i$ is not zero.
This decision makes this analysis an over-approximation.
The other option would be to mark the variable as zero when it is marked as zero on both of the branches.
This analysis marks none of the variables as zero in \cref{handout:div_zero}.
There is a false negative in the last line of the code snippet in this case, this makes the analysis an under-approximation.

Obviously, there is room for improvement which is called path-sensitive analysis.
In path-sensitive analysis, each possible execution path of the program is considered.
This method this method can achieve a great precision.
The analysis can exclude some impossible branches and it can track the possible values of the variables more precisely.
There is a cost to this precision in execution time, as this kind of analysis analysis has exponential complexity in the number of branches.
Fortunately, the run-time of such analysis turns out to be manageable in practice.
There are existing tools that use this method on industrial scale code-base.
In case the analyser tracks the possible values of variables as symbols and does symbolic computation with them, we are talking about symbolic execution.
To track the possible values of those symbols a constraint manager and a constraint solver are used.
One of the main sources of those  constraints are conditions on the branches of the CFG.

Symbolic execution is a way of performing abstract interpretation.
Abstract interpretation is a way of interpreting programs, but instead of using the concrete semantics (the most precise interpretation) of the program we use an abstract semantics.
The reason of defining abstract semantics is to make the problem more tractable.

Software semantics are not always determined by the program code.
It can also depend on compilation parameters.
For this reason, it is crucial to know those parameters during abstract interpretation.
Those parameters can be macro definitions, the language standard choice, platform etc.
This makes determining such parameters part of the static analysis in most of the cases.
Most of the time in practice this is achieved by the logging of the arguments passed to the compiler during a compilation.

Note that there are several formal methods -- a lot of them are based on abstract interpretation -- that are used to verify the correctness of software, but they are out of the scope of this paper.
It focuses on practical tools with large-scale industrial applications.

Once we have flow-sensitive or path-sensitive analysis it is interesting to consider what is the biggest scope that the analyser can reason about using these approaches.

\subsection{Dynamic analysis}

In dynamic analysis, the run-time behaviour of the program is analysed by special code compiled into the binary, sometimes with the aid of a run-time library or a virtual execution environment.

Dynamic analysis incurs a run-time cost, but it can be more precise than static source code analysis. Since we have knowledge about the execution environment, it is possible to achieve fewer false positives.
On the other hand, only executed code can be analysed.
In case a bug is not triggered or a piece of code is not covered, we will miss the problem.
Static analysis has the possibility to discover errors in uncovered code or find unexpected errors that are not checked by our tests.

Also, there are some properties of the code that cannot be tested or analysed dynamically.
These properties include naming conventions, code formatting, misusing some language features, code smells, and portability issues.

LLVM has a number of dynamic analysis tools, the sanitisers.
These can find memory addressing issues, memory leaks, race conditions, and other sources of undefined behaviour like division by zero.
There are other popular tools outside of the LLVM project such as Valgrind~\cite{10.1145/1273442.1250746}.

We do not cover the details of dynamic analysis in this paper but it is a useful tool to have at your disposal.
Static and dynamic analysis are two different tools for slightly different purposes.
It is better to use both techniques to get the best of both worlds.

\section{A quick tutorial on CodeChecker}\label{codechecker}

CodeChecker~\cite{CodeChecker} is a static analysis infrastructure tooling system originally built on top of the LLVM/Clang Static Analyzer and Clang-Tidy software.
The initial goal was to offer a replacement for \texttt{scan-build} in a Linux or macOS (OS/X) development environment.
\texttt{scan-build} is a script supplied with the LLVM/Clang Compiler Infrastructure that allows users to run the Clang Static Analyzer on their project.
CodeChecker was developed to allow greater customisation of the executed build, to support Clang-Tidy, and to support the viewing of the analysis results in a controlled, centralised, remotely accessible web application.
Ever since its inception, CodeChecker has grown to be a feature-rich software defect analysis and triaging system.

The main features of CodeChecker are as follows.

\begin{itemize}
    \item Support for executing multiple analysers on your project (currently the Clang Static Analyzer and Clang-Tidy).
    \item Fine-tuning of analysis modules' at the invocation of analysis, without having to meddle with the individual analysers.
    \item Subsequent analysis runs only check and update results for modified files without analysing the entire project (depends on build system support!)
    \item Suppression of known false positive results, either using a configuration file or via annotation in source code, along with the exclusion of entire source paths from analysis
    \item Support storage of analysis reports from (but not the execution of) a plethora of static analysers, including Java, Python, and Go analysers.
    \item Web application which allows viewing and discussing discovered code defects, with a streamlined and easy experience.
    \item Filtering of results across almost all parameters.
    \item Comparison (diff) view to contrast individual analysis runs, e.g.\ to gatekeep defect-introducing changes.
    \item Self-contained static HTML report generation in case the full Web application experience is not needed.
    \item Easily implementable Thrift-based server-client communication used for the storage and query of discovered defects
    \item Support for multiple bug visualisation front-ends, such as the web application, a command-line tool, and an Eclipse plug-in
\end{itemize}

\subsection{Quick tutorial}
The commands and workflow explained in this paper is tailored for the latest release of CodeChecker available at the time of writing, version \texttt{6.14.0}.
CodeChecker can be downloaded from the official repository on GitHub~\cite{CodeChecker}.
After installation of a few system packages, CodeChecker can be built, and the resulting script's location should be added to the \texttt{PATH} environment variable.

Afterwards, navigate to your project's build directory, and you can analyse your project by doing \emph{a full build} through the CodeChecker \texttt{check} command.
The invocation and the output of \texttt{check} is presented in \cref{handout:codechecker:check}.
The individual reports' main diagnostic message, and a summary table on the numbers of analysis rules (``checkers'') that reported is output immediately.

\begin{listing}[H]
    \begin{minted}{Bash}
        $ CodeChecker check --build "make clean; make" --output ./reports
    \end{minted}
    \begin{minted}{Text}
        [INFO] - Starting static analysis @\ldots@
        @\color{darkgray}{\emph{\ldots}}@
        [HIGH] main.cpp:13:14: Division by zero [core.DivideZero]
        return 1 / zero();
                 ^
        Found 1 defect(s) in main.cpp

        ----==== Summary ====----
        Filename | Report count
        -----------------------
        main.cpp |            1

        Severity | Report count
        -----------------------
        HIGH     |            1

        Total number of reports: 1
    \end{minted}
    \caption{The invocation and output of the \texttt{CodeChecker check} command.}\label{handout:codechecker:check}
\end{listing}

To view the results in the web application, you need to start a server by executing \mintinline{Bash}{CodeChecker server}.
Once the server is running, the \mintinline[breaklines]{Bash}{CodeChecker store ./reports -n my-project} command will store the results in the \texttt{check} command's output directory to the server.
The server is available on the local machine, by navigating to \url{http://localhost:8001} with a web browser. A screenshot of the CodeChecker web application can be seen in Figure 4. Figure 5 shows a visualisation of a path-sensitive bug report.

\begin{sidewaysfigure}
    \centering
    \includegraphics[width=\textwidth]{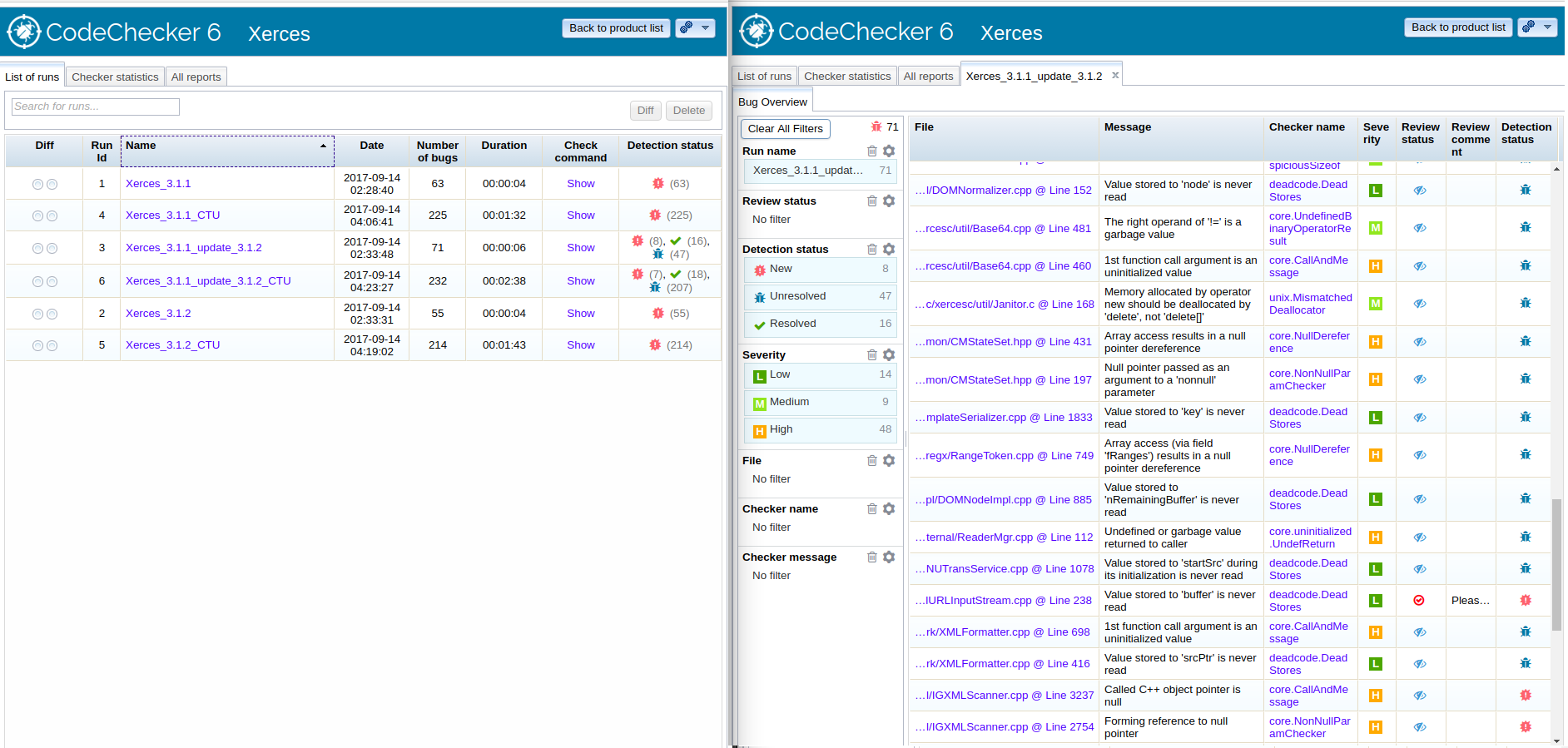}
    \caption{A screenshot of the CodeChecker web application. The left-hand side shows a list of individual analysis invocations (\emph{runs}) on the open-source Xerces project, while the right-hand side lists the individual bug reports from the analysers.}\label{handout:BugList}
\end{sidewaysfigure}

\begin{sidewaysfigure}
    \centering
    \includegraphics[width=\textwidth]{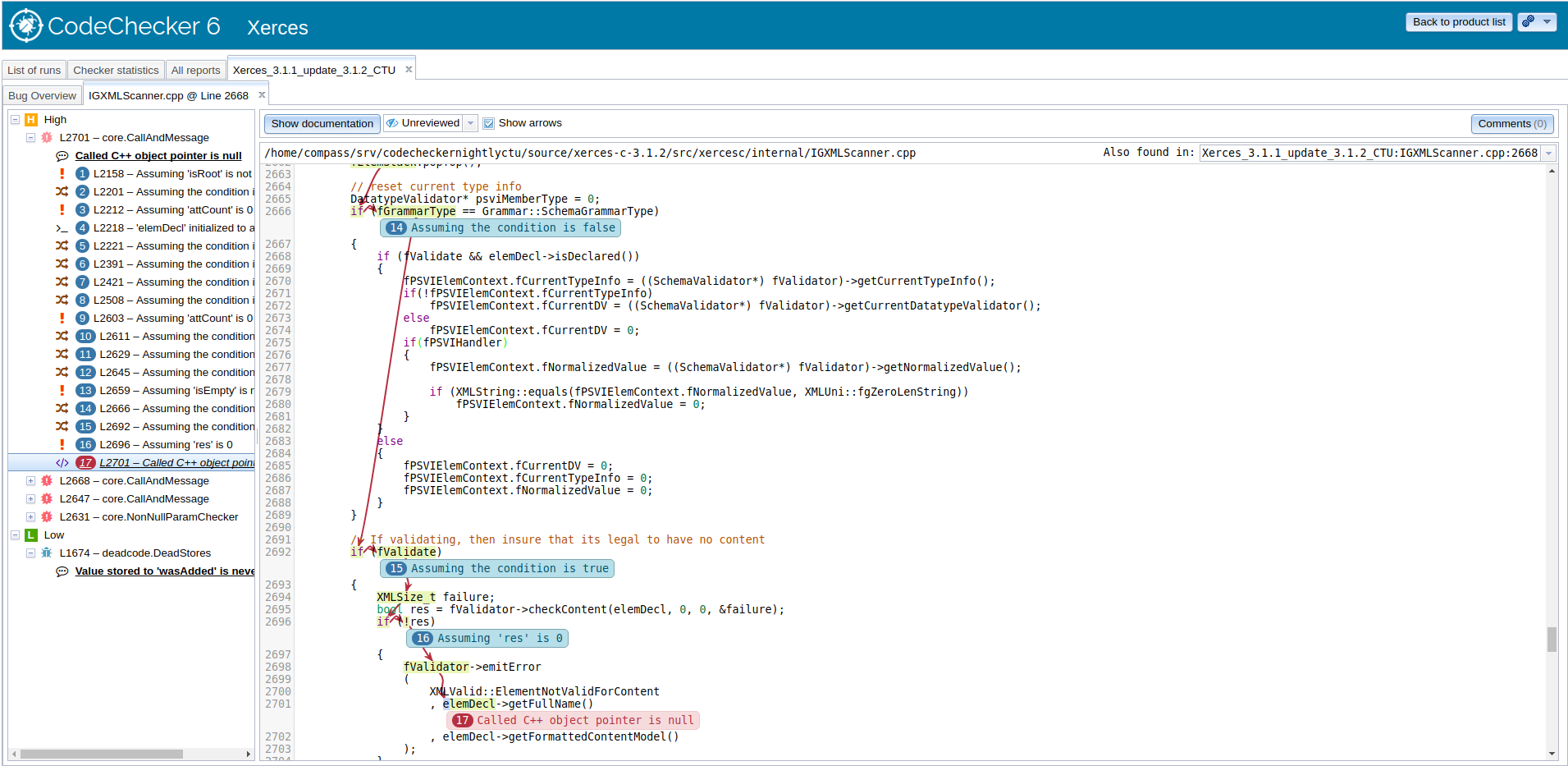}
    \caption{Visualisation of a Clang Static Analyzer \emph{bug path} -- the steps the analyser took at ``executing'' the program which resulted in the report.}\label{handout:BugPath}
\end{sidewaysfigure}

\clearpage
\section{Syntax Tree-based Analysis}\label{tidy}

The easiest way of writing an analysis routine that matches a particular piece of code would be using textual matching and regular expressions against tokens in the source code (see \cref{handout:methods}).
However, such approaches have serious issues when applied for C and \CC{} source code for a variety of reasons.
First, macros allow the compiled source code to be different from the source code that's physically written in the file.
Second, the \mintinline{CPP}{typedef} and \mintinline{CPP}{using} statements allow the creation of type aliases for any type, and thus, a rule that is defined for the underlying type (e.g.\ \mintinline{CPP}{int}) will not match through the created aliases (e.g.\ \mintinline{CPP}{MyInt}), even though the real type and the compiled code should match the rule.
In addition, the \CC{} grammar is not regular, and there are various ambiguities that can only be resolved in a context-sensitive fashion through type information.
Due to these problems, it is more robust to use a type-safe and proper representation of the source code built by a compiler, such as an \emph{abstract syntax tree} (see \cref{handout:parsing}).
This approach is generally cheap enough to run even during active development of the source code, e.g.\ through an editor.

\begin{listing}
    \begin{sublisting}[t]{.5\textwidth}
        \begin{minted}[bgcolor={}]{CPP}
            Something* p = function_call();
            int value = p@$\rightarrow$@value;
            std@$::$@cout << value;
        \end{minted}
        \caption{Original code.}\label{tidy:redundant-ptr/original}
    \end{sublisting}\hfill%
    \begin{sublisting}[t]{.5\textwidth}
        \begin{minted}[bgcolor={}]{CPP}
            int value = function_call()@$\rightarrow$@value;
            std@$::$@cout << value;
        \end{minted}
        \caption{After elimination of \mintinline{CPP}{p}.
            This rewrite is possible in any \CC{} standard version.}\label{tidy:redundant-ptr/rewrite}
    \end{sublisting}
    \caption{Example code where the pointer variable \mintinline{CPP}{p} is superfluous, as it is only used once.
        We can also infer that the developer ``knows'' the value can not be a \emph{null pointer}, as no conditions guard the dereference.}\label{tidy:redundant-ptr}
\end{listing}

In this section, we will discuss the design and implementation of a syntax-tree-based analysis rule that targets the elimination of \emph{redundant pointer variables} from the program code.
The details of the research behind this refactoring and the findings of such cases on real-world projects can be found in~\cite{Szalay2020Towards}.
A redundant variable causes problems during code comprehension and development, mostly through polluting the set of available variables and causing barriers in refactoring routines.
We can eliminate the occurrences of such variables by rewriting the code.
Because this operation is a modernisation refactoring, we have to ensure that the code's semantics, and thus the program's behaviour does not change.
Due to the issues as mentioned earlier, it may not be trivial to deduce that a variable is a pointer from merely looking at the tokens.
As such, syntax-tree based analysis is needed.

\begin{listing}[h!]
    \begin{sublisting}[t]{\textwidth}
        \begin{minted}[bgcolor={}]{CPP}
            Something* p = function_call_that_might_return_null();
            if (!p)
                return; // Pointer is null. Dereference here is undefined!

            int value_to_print = p@$\rightarrow$@value;
            std@$::$@cout << "Returned value: " << value_to_print;
            // p in scope here. (This is what we want to eliminate!)
        \end{minted}
        \caption{Original code.}\label{tidy:null-check/original}
    \end{sublisting}\hfill%
    \begin{sublisting}[t]{\textwidth}
        \hspace*{1.5cm}
        \begin{minted}[bgcolor={}]{CPP}
            int value_to_print;
            if (Something* p = function_call_that_might_return_null(); !p ||
                    ((value_to_print = p@$\rightarrow$@value), false))
                return;

            // p not in scope here.
            std@$::$@cout << "Returned value: " << value_to_print;
        \end{minted}
        \caption{The rewrite of the code where the scope of \mintinline{CPP}{p} is limited to the necessary checks.
            The initialisation in the conditional is possible in \CC{}17 and newer standards.}\label{tidy:null-check/rewrite}
    \end{sublisting}
    \caption{Example code where the \mintinline{CPP}{p} pointer variable is used only once in a dereference that initialises another variable, in addition with a guard against its null value.
        The rewrite must ensure that the semantics are kept, and no null dereference occurs.}\label{tidy:null-check}
\end{listing}

\clearpage 
We can categorise the pointer-type variables into the following categories:

\begin{itemize}
    \item \textbf{Unused variables.}
        We do not need to consider this.
        Built-in compiler warnings exist to warn and eliminate unused variables.
    \item Pointer variable \textbf{used only once}.
        \cref{tidy:redundant-ptr} shows such an example where the pointer is only used in a dereference, once.
        By inlining the expression that the pointer was initialised with, the variable can be eliminated.
    \item Pointer variable \textbf{dereferenced} only once to initialise another variable, and \textbf{\textit{guarded} in one location}.
        This case is less trivial than the previous.
        The example of this case can be seen in \cref{tidy:null-check}.
        Because dereferencing a \emph{null pointer} is an operation that results in undefined behaviour in \CC{}, developers regularly use a construct like depicted.
        A \emph{guard operation} is defined to be a conditional branch on a simple arithmetic check of pointer's value\footnote{%
            This is most commonly done by checking for or against \emph{null pointers}, but additional checks, such as the pointer being in the range of an array, are applicable.
        } followed by a simple control flow breaking statement\footnote{%
            In most cases, a structure of \emph{early \mintinline{CPP}{return}s} is favoured.
            Similar constructs, such as \mintinline{CPP}{break} or \mintinline{CPP}{continue} in loops, the \mintinline{CPP}{throw}ing of an exception are also considered ``flow-breaking''.
            Besides, every \emph{non-returning} function which are attributed with \mintinline{CPP}{@$[[$@noreturn@$]]$@}, terminating the program (such as \mintinline{CPP}{std@$::$@abort}) are also ``flow breaking''.
        } that terminates the execution before a dereference of an invalid pointer could take place.
        We will try to move the pointer into an inner scope and perform the \emph{guard} operation, while the dereference's result variable lives in the outer scope.
    \item Pointer variable \textbf{used multiple times}, guarded multiple times, the guard statement is complex, etc.
        In this case, the dedicated pointer variable is justified, and we do not rewrite the code.
\end{itemize}

For the transformation in \cref{tidy:null-check/rewrite} to be valid, a few additional conditions must be satisfied.
Consider how the result variable is declared (\mintinline[breaklines,breakafter=_]{CPP}{int value_to_print;}) and how the assignment after the guard, involving the dereference, occurs (\mintinline{CPP}{value_to_print = p@$\rightarrow$@value;}).
The rewriting of the code is only possible if the type of the initialised variable supports these operations: default construction and assignment.
In addition, the project under analysis by the tool has to be compiled with at least \CC{}17 standard version, as the creation of a local variable in a conditional branch and having a condition is only supported since the release of this standard.
If these conditions are satisfied, we can sufficiently rewrite the code.
The formal skeleton of the rewrite is presented in \cref{tidy:deref-init-rewrite}.
There are intricacies to the rewrite that might not be apparent at first glance, and thus, in \cref{tidy:deref-truth} we explain the semantics of the operations.

\begin{listing}
    \begin{sublisting}[t]{.49\textwidth}
        \begin{minted}[bgcolor={}]{CPP}
            T* p = @$\Pi$@();
            if (@$\phi$@(p))
                return;
            V v = @$\lambda$@(p);
        \end{minted}
        \caption{Skeleton of the original code, with details removed.}\label{tidy:deref-init-rewrite/original}
    \end{sublisting}\hfill%
    \begin{sublisting}[t]{.49\textwidth}
        \begin{minted}[bgcolor={}]{CPP}
            V v;
            if (T* p = @$\Pi$@(); (!@$\phi$@(p) ||
                    ((v = @$\lambda$@(p)), false)))
                return;
        \end{minted}
        \caption{The formalised version of the rewrite of this example.}\label{tidy:deref-init-rewrite/rewrite}
    \end{sublisting}
    \caption{Formal overview of the rewrite of a pointer that is only used once in a dereference that initialises another variable, with a \emph{guard ($\phi$) branch}.
        \mintinline{CPP}{@$\lambda$@(p)} refers to the dereferencing usage of \mintinline{CPP}{p} in the initialiser assignment.}\label{tidy:deref-init-rewrite}
\end{listing}

\begin{table}[]
\centering
\caption{The truth table for the semantics of the example in \cref{tidy:deref-init-rewrite}. For clarity, we assume that \mintinline{CPP}{@$\phi$@(p)} is the null pointer dereference guard, \mintinline{CPP}{!p} (in longer form: \mintinline{CPP}{p != nullptr}).}
\label{tidy:deref-truth}
\resizebox{\textwidth}{!}{%
\begin{tabular}{ccc|cccccc}
\multicolumn{3}{c|}{\textit{Original code}} &
  \multicolumn{6}{c}{\textit{Rewritten code}} \\ \hline
\textbf{\mintinline{CPP}{p}} &
  \textbf{\mintinline{CPP}{@$\phi$@(p)}} &
  \textbf{\begin{tabular}[c]{@{}c@{}}\mintinline{CPP}{return}\\ taken\end{tabular}} &
  \textbf{\mintinline{CPP}{p}} &
  \textbf{\mintinline{CPP}{@$\phi$@(p)}} &
  \textbf{\begin{tabular}[c]{@{}c@{}}assignment\\ takes place\end{tabular}} &
  \textbf{\begin{tabular}[c]{@{}c@{}}value of\\ \mintinline{CPP}{(asg, false)}\end{tabular}} &
  \textbf{\begin{tabular}[c]{@{}c@{}}value of full \\ condition (\mintinline{CPP}{||})\end{tabular}} &
  \textbf{\begin{tabular}[c]{@{}c@{}}\mintinline{CPP}{return}\\ taken\end{tabular}} \\ \hline
\textit{not null} &
  \cross &
  \cross &
  \textit{not null} &
  \cross &
  \tick &
  \mintinline{CPP}{false} &
  \mintinline{CPP}{false} &
  \cross \\
\textit{\mintinline{CPP}{nullptr}} &
  \tick &
  \tick &
  \textit{\mintinline{CPP}{nullptr}} &
  \tick &
  \cross &
  not evaluated as \mintinline{CPP}{||} is short-circuiting &
  \mintinline{CPP}{true} &
  \tick
\end{tabular}%
}
\end{table}

\subsection{The \emph{Abstract Syntax Tree} (AST)}\label{tidy:ast}
The \emph{Abstract Syntax Tree} data structure is the core representation of the parsed source code in Clang.
This view encodes not only the source text's contents like identifiers but the necessary structural information, too.
The ambiguities of the \CC{} language come from tokens only resolved through type information -- such as the case of token sequence \mintinline{CPP}{T * p} being either a multiplication (or, rather, an \mintinline{CPP}{operator *} call) or a pointer variable based on what \mintinline{CPP}{T} is -- the AST necessarily contains the type information.
An example of three different resolutions of the same token sequence can be seen in \cref{tidy:ambiguous-ast}.
The AST is the input for various compile-time analyses\footnote{%
    E.g.\ control-flow analysis, data-flow analysis, dead code elimination, etc.
} and the code generator logic.
While the internal implementation of the AST are heavily optimised~\cite{Carruth2016High}, the library offers users an object-oriented hierarchy of types that represent elements of the \CC{} grammar and the AST is built as a pointer-linked data structure of such instances.
The three major distinct hierarchies are \mintinline{CPP}{Decl} for declarations and scope contexts, \mintinline{CPP}{Stmt} (and its subclass, \mintinline{CPP}{Expr}) for statements and expressions, and \mintinline{CPP}{Type} for the representation of type details.

\begin{figure}
    \centering
    \includegraphics[height=.4\textheight, keepaspectratio]{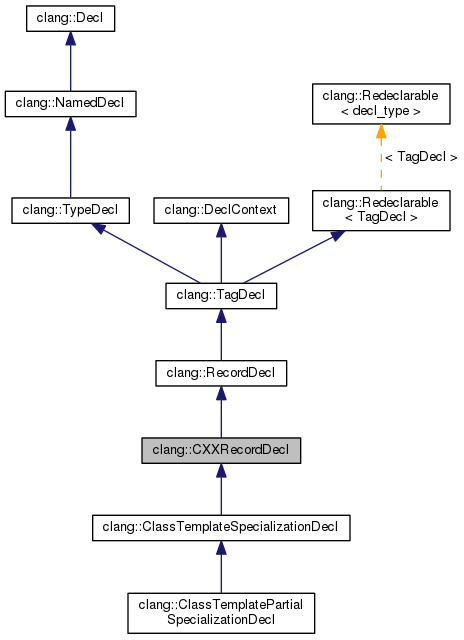}
    \caption{Inheritance diagram for \mintinline{CPP}{clang@$::$@CXXRecordDecl}, the AST node type for \mintinline{CPP}{class}es and other record types.}\label{tidy:cxxrecorddecl}
\end{figure}

\begin{figure}
    \begin{subfigure}{\textwidth}
        \includegraphics[width=\linewidth]{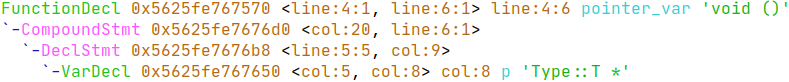}
        \caption{Declaring a pointer variable (\mintinline{CPP}{VarDecl}) \mintinline{CPP}{p} to a pointee of type \mintinline{CPP}{T}.}\label{tidy:ambiguous-ast/pointer-decl}
    \end{subfigure}\hfill%
    \begin{subfigure}{\textwidth}
        \includegraphics[width=\linewidth]{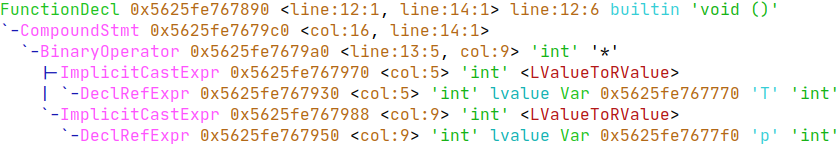}
        \caption{Multiplying two \mintinline{CPP}{int} variables, \mintinline{CPP}{T} and \mintinline{CPP}{p}.
            The built-in $\star$ operator (\mintinline{CPP}{BinaryOperator}) is called.}\label{tidy:ambiguous-ast/builtin-multiply}
    \end{subfigure}\hfill%
    \begin{subfigure}{\textwidth}
        \includegraphics[width=\linewidth]{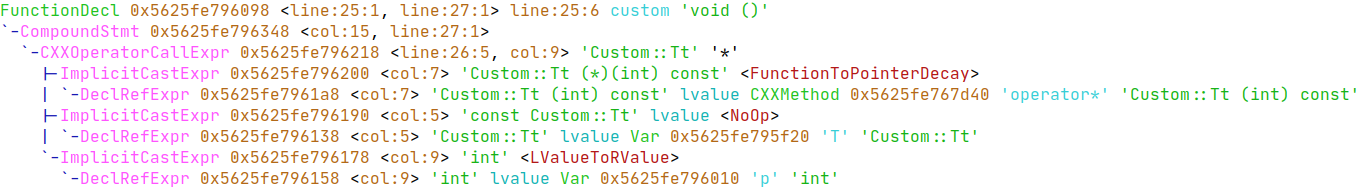}
        \caption{Performing ``multiplication'' of a user-defined \mintinline{CPP}{T} (e.g.\ \emph{Matrix}) and an \mintinline{CPP}{int p}.
            As \mintinline{CPP}{T} has the overloaded \mintinline{CPP}{operator *}, that function is called (\mintinline{CPP}{CXXOperatorCallExpr}).}\label{tidy:ambiguous-ast/custom-operatorstar}
    \end{subfigure}
    \caption{The \CC{} grammar is ambiguous.
        For example, the token sequence \mintinline{CPP}{T * p;} may be understood in multiple ways, depending on the types of \mintinline{CPP}{T} and \mintinline{CPP}{p}.
        The \emph{AST} represents the parsed syntax, where such ambiguities had been resolved.}\label{tidy:ambiguous-ast}
\end{figure}

Clang's \emph{AST} offers an object-oriented representation complete with an elaborate type hierarchy to represent the syntax of parsed \CC{} sources.
\cref{tidy:cxxrecorddecl} contains the inheritance graph for \mintinline{CPP}{clang@$::$@CXXRecordDecl} as generated by \emph{Doxygen} for LLVM/Clang $10.0$.
The inheritance relations express categories and capabilities of nodes robustly: we can see, that the \CC{} record declarations are declarations, named declarations, and define types in the language.
In addition, \CC{} records are declaration contexts, i.e.\ they may contain declarations in themselves, such as fields and member functions, and that they are \emph{redeclarable}, which means that multiple declarations may exist for these kinds of nodes.

\subsection{Matching Nodes and Trees in the AST}\label{tidy:matching}

The \mintinline{CPP}{ASTMatcher} library is part of the Clang suite and allows writing matcher expressions via a declarative syntax, through a general-purpose library employing template metaprogramming.
The matcher expression that finds every variables named $x$ is expressed as \mintinline{CPP}{varDecl(hasName("x"))}.
This matcher is executed on a sub-tree via a simple function call, \mintinline[breaklines]{CPP}{ast_matchers@$::$@match(Matcher, Node, Node@$\rightarrow$@getASTContext());} which will return us the set of nodes matched.
The library contains numerous matchers, which can be combined with special combinators and adaptors.
There are four main combinators between matchers: embedding a matcher into another one (\mintinline{CPP}{varDecl(hasName("x"))}) filters the result of the outer matcher by the inner predicate.
\mintinline{CPP}{anyOf()} and \mintinline{CPP}{allOf()} link sub-matchers as siblings, expressing the $\cup$ and $\cap$ set operations, while \mintinline{CPP}{unless()} expresses $\neg$, making the matcher not match if the inner predicate matches.
In addition to these, there are traversal matchers, such as \mintinline{CPP}{hasDescendant()}, \mintinline{CPP}{hasParent()}, etc.\ which can be used to walk and filter the tree directly, if needed.
From these simple constructs, various complex expressions can be built.
In the software code, the users can also save matcher expressions into variables and reuse them without duplication, which allows for more readable client code.
All node matchers support a \mintinline{CPP}{bind()} operation, which assigns a name to the bound node.

There are two main clients in the Clang suite for the AST matcher library: \texttt{clang-query} and \texttt{clang-tidy}.
Clang-Query is a command-line REPL\footnote{%
    \emph{REPL}: \emph{R}ead-\emph{E}valuate-\emph{P}rint \emph{L}oop.
} tool in which the users might write the aforementioned matcher expressions, and the tool prints the nodes that are matched, highlighting the source code fragments where the nodes were parsed from.
Clang-Query is more intended for developers of the library as it can not offer custom handling of the matches -- there is only one handler, ``print''.

Clang-Tidy is a user-facing tool and a framework which automates the execution of checks.
Checks are organised into groups,\footnote{%
    Groups are formed based on the general idea (\texttt{modernise}, \texttt{performance}, etc.) or guideline (\texttt{cppcoreguidelines}, \texttt{linux} kernel, etc.) the checks target.
} and the user can specify which set of checks to run for the source file.
At the time of writing of this paper, Clang-Tidy contains $239$ unique checks.
In addition, Clang-Tidy offers an easy to use template for checks, automating the execution of matchers and the callback with results to the check code.
We have implemented the aforementioned rule in Clang-Tidy.
The following sections in this paper were written with LLVM/Clang version $10.0$ in mind.

Earlier, G\'{a}bor Horv\'{a}th developed \emph{CppQuery}~\cite{Horvath2014Szemantikus,CppQuery} as graphical tool which aids the combination of arbitrary matcher expressions and seeing their result.

\subsection{Skeleton of a Clang-Tidy check implementation}
We will name our check \emph{Redundant Pointer in Local Scope} and put it into the \texttt{readability} group.
Clang-Tidy checks are implemented as \CC{} classes inheriting from \mintinline{CPP}{ClangTidyCheck}.
Conventionally, the name of the check is included in the class's name, and the name of the files related to the implementation.
Thus, we will create the two files depicted in \cref{tidy:check-skel.h,tidy:check-skel.cpp}.

\begin{listing}
\begin{minted}[bgcolor={}]{CPP}
#ifndef READABILITY_REDUNDANTPOINTERCHECK_H
#define READABILITY_REDUNDANTPOINTERCHECK_H
#include "../ClangTidyCheck.h"

namespace clang {
namespace tidy {
namespace readability {

class RedundantPointerCheck : public ClangTidyCheck {
public:
  RedundantPointerCheck(StringRef Name, ClangTidyContext* Context)
    : ClangTidyCheck(Name, Context) {}
  void registerMatchers(ast_matchers@$::$@MatchFinder* Finder) override;
  void check(const ast_matchers@$::$@MatchFinder@$::$@MatchResult& Result)
    override;
};

} // namespace readability
} // namespace tidy
} // namespace clang

#endif // READABILITY_REDUNDANTPOINTERCHECK_H
\end{minted}
\caption{The class definition for our new \emph{redundant pointer variable} check in the header file \texttt{readability/RedundantPointerCheck.h}.}\label{tidy:check-skel.h}
\end{listing}

\begin{listing}
\begin{minted}[bgcolor={}]{CPP}
#include "RedundantPointerCheck.h"
#include "clang/AST/ASTContext.h"
#include "clang/ASTMatchers/ASTMatchFinder.h"
using namespace clang@$::$@ast_matchers;

namespace clang {
namespace tidy {
namespace readability {

void RedundantPointerCheck@$::$@registerMatchers(MatchFinder* Finder)
  { /* $\ldots$ */ }
void RedundantPointerCheck@$::$@check(const MatchFinder@$::$@MatchResult& Result)
  { /* $\ldots$ */ }

} // namespace readability
} // namespace tidy
} // namespace clang
\end{minted}
\caption{The empty implementation file that defines the necessary methods for our new \emph{redundant pointer variable} check in the source file \texttt{readability/RedundantPointerCheck.cpp}.}\label{tidy:check-skel.cpp}
\end{listing}

To add the check to the list of source files that need to be built with Clang-Tidy, we need to add the implementation source file \texttt{RedundantPointerCheck.cpp} to list of source files in \texttt{readability/CMakeLists.txt}.
This will ensure that the new check's code is compiled when Clang-Tidy is next built.
However, we also need to give a ``user-facing'' name to the check.
Every check group in Clang-Tidy is organised into a module, in this case, described in the \CC{} source file \texttt{ReadabilityTidyModule.cpp}.
This file needs to be extended with the inclusion of the new check's header: \mintinline{CPP}{#include "RedundantPointerCheck.cpp"}, and in the \mintinline{CPP}{addCheckFactories()} function, calling \mintinline[breaklines,breakanywhere]{CPP}{CheckFactories.registerCheck<RedundantPointerCheck>("readability-redundant-pointer");}.
After taking all these steps, the new, empty check is wired into the rest of the framework.

\newcommand{\callM}{\mathcal{M}}

The check's body is comprised of two important functions, \linebreak
    \mintinline{CPP}{registerMatchers()} and \mintinline{CPP}{check()}.
\mintinline{CPP}{registerMatchers()} is run automatically by Clang-Tidy for each check, and the responsibility of this function is to set up the matcher expressions that will call back the check's implementation.
The convention is to use a series of \mintinline{CPP}{Finder@$\rightarrow$@addMatcher(@$\callM$@, this)} calls, where $\callM$ is the matcher expression.
Note that the order of matchers' registration is not indicative of the order of the callbacks firing!
When a matcher registered by the current check implementation matches, the \mintinline{CPP}{check()} callback is fired, this is why the \mintinline{CPP}{this} argument is passed to the \mintinline{CPP}{addMatcher()} function.

\let\callM\undefined

\subsection{Matching the necessary nodes and code parts}\label{tidy:matcher-exprs}
The options for creating matcher expressions are vast, as there exists, in general, one matcher expression primitive for each type of node in the AST.
Matcher expressions can be combined and reused.
The \emph{AST Matcher} library uses a combination of template metaprogramming and dynamic dispatch techniques to work on the inside and offers a concise declarative syntax on the outside.
The matchers are constructed by calling a factory function that takes a set number of arguments, usually sub-matchers or property expressions, such as \mintinline{CPP}{varDecl(hasName("x"), hasType(cxxRecordDecl()))}.
This matcher will fire for every \mintinline{CPP}{VarDecl} node in the AST with the variable name \mintinline{CPP}{x} and which declared type is any \mintinline{CPP}{CXXRecordDecl}.
The \CC{} record types are the union of the \mintinline{CPP}{class}, \mintinline{CPP}{struct}, and \mintinline{CPP}{union} types.
During the development of Tidy checks, it is advised to continuously consult both the \emph{AST Matcher Reference Guide}~\cite{ASTMatchers} and the documentation for AST node types.

The matchers constructed by the calls to these factory functions are, however, valid instantiated \CC{} objects, and as such, can be bound to variables.
\mintinline{CPP}{static const auto Submatcher = @$\ldots$@;} followed by a call to \linebreak
    \mintinline[breaklines]{CPP}{Finder@$\rightarrow$@addMatcher(someOuterMatcher(Submatcher), this);}
is a common idiom for organising matchers into reusable components, a technique we also employ.

\begin{figure}
    \centering
    \includegraphics[width=\textwidth]{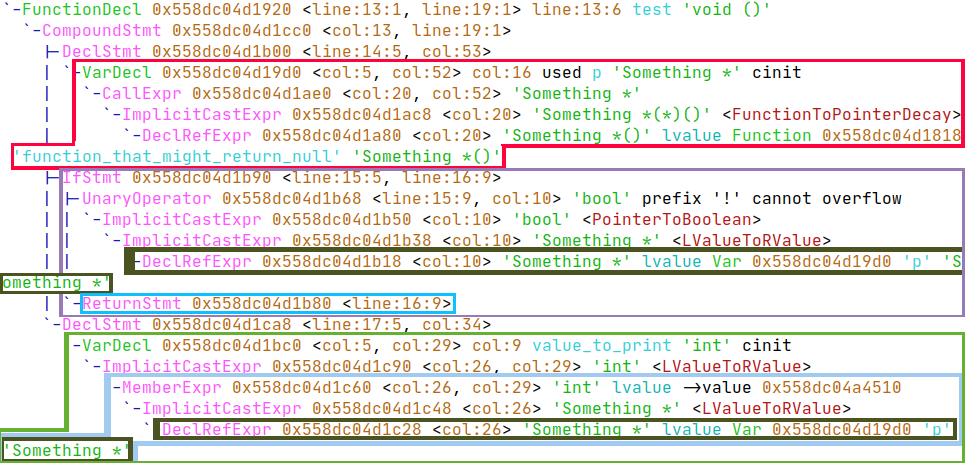}
    \caption{The example of \cref{tidy:deref-init-rewrite/original} represented as a Clang Abstract Syntax Tree (see \cref{tidy:ast}).
        The coloured boxes highlight the matchers we implemented, selecting sub-trees.
        The legend for the individual colours and their matches:
        \colorbox{americanrose}{\color{white} \texttt{PointerVar}},
        \colorbox{armygreen}{\color{white} \texttt{VarUsage}} (see \cref{tidy:ptr-var});
        \colorbox{babyblueeyes}{\texttt{Dereference}} (see \cref{tidy:ptr-derefs});
        \colorbox{green(ryb)}{\texttt{VarInitFromDerefence}} (see \cref{tidy:init-matcher});
        \colorbox{spirodiscoball}{\texttt{FlowBreakingStmt}},
        \colorbox{lavenderpurple}{\texttt{Guard}} (see \cref{tidy:guard-stmt})
        }\label{tidy:matchers}
\end{figure}

\subsubsection{Matching pointer variables and their simple usage points}\label{tidy:ptr-var}
Let us now think back to the goals of our \emph{redundant pointer variable check} (see \cref{tidy}), and identify individual sub-trees in the AST that are interesting to us.
First, we need to get the \textbf{pointer variables} in the analysed code, which have an initialiser at their declaration.
This is done similarly to the previous example: \mintinline[breaklines]{CPP}{PointerVar = varDecl(hasType(pointerType()), hasInitializer(expr()))}.
In \linebreak addition, we need \textbf{every usage} of these variables.
A reference of a declaration is captured in the AST as a \mintinline{CPP}{DeclRefExpr}, from which we can narrow down to usages of declarations that match the previous matcher: \mintinline[breaklines]{CPP}{VarUsage = @$\linebreak$@ declRefExpr(to(PointerVar))}.
An important member function that can be called on matcher objects is \mintinline{CPP}{bind()}: bind allows to assign a string identifier to the matcher's (or in case of a complicated match expression, the submatchers') result.
As we are registering multiple matchers and reusing the same sub-matchers in multiple contexts, it will be necessary to distinguish the individual results, and thus we will use \mintinline{CPP}{bind} abundantly.

In a complex expression (such as \mintinline{CPP}{int val = ptr@$\rightarrow$@x;}) the \texttt{VarUsage} is only the ``innermost'' sub-tree, namely, the tree for the typed code fragment \mintinline{CPP}{ptr}.
In \cref{tidy}, we explained that we want to separately target two cases: unique usages (without initialisation), and unique usages for initialisation of another variable (potentially guarded).
These require additional matchers.
During the development of syntax-tree based checks, continuous consideration through test-driven development and checking of how the test code samples' are represented as syntax trees are needed.
The example in \cref{tidy:deref-init-rewrite/original} shows the most complex case, which we show the AST, highlighting how the individual matchers match this structure in \cref{tidy:matchers}.

\subsubsection{Matching dereferences of pointers}
\begin{listing}
    \begin{minted}[bgcolor={}]{CPP}
        static const auto Dereference = stmt( // Every statement, for which $\ldots$
            anyOf( // $\ldots$ either of the following "predicates" is true $\ldots$
                // a) It is dereferenced for a data member.
                VarMemberUsage.bind("DerefUsage"),

                // b) It is dereferenced for a function call or call of member.
                cxxMemberCallExpr(has(VarMemberUsage)).bind("DerefUsage"),

                // c) It is a plain simple dereference:
                unaryOperator(
                    hasOperatorName("*"), // The application of the $\star$ operator $\ldots$
                    hasDescendant(VarUsage.bind("DerefdVar")) // $\ldots$ to a pointer.
                ).bind("DerefUsage")
            ));
    \end{minted}
    \caption{The matcher expression that captures the contexts where a pointer variable is dereferenced.}\label{tidy:ptr-derefs}
\end{listing}

Dereference of a pointer can take three forms: plain dereference (\mintinline{CPP}{*p}), data member access (\mintinline{CPP}{p@$\rightarrow$@x}) and member function call (\mintinline{CPP}{p->f()}).\footnote{%
    The ``long form'' of the $\rightarrow$-notation, \mintinline{CPP}{(*p).x} is represented in the syntax tree with the same nodes, with the \mintinline{CPP}{isArrow()} query telling the client if $\rightarrow$-notation is used.
}
First, we will create the matcher that matches the latter cases, as this is special from our check's point of view.
\mintinline[breaklines]{CPP}{VarMemberUsage = memberExpr(hasDescendant(VarUsage.bind("DerefdVar")))} captures the member references through usage of a pointer variable.
We can wrap this matcher to account for member functions' calls, by creating the \linebreak
    \mintinline{CPP}{cxxMemberCallExpr(has(VarMemberUsage))}\footnote{%
    In the \CC{} grammar, member function calls are calls of members, the notion of ``member function'' does not exist during parsing of the ``member select expression''.
    Member calls are actually parenthesised as \mintinline{CPP}{(x@$\rightarrow$@f)()}, but the left-hand parentheses are implicit.
    The expression is read as \emph{``select the member \texttt{f} from \texttt{x} and perform a call on it''}.
} matcher.
The full set of dereferences can be expressed as yet another matcher built from the previous ones, as depicted in \cref{tidy:ptr-derefs}.

\subsubsection{Matching initialisation of variables (from a dereference)}
\begin{listing}[h]
    \begin{minted}[bgcolor={}]{CPP}
        static const auto ConstructionExprWithDereference =
          ignoringElidableConstructorCall(
            cxxConstructExpr(       // Constructor calls $\ldots$
              argumentCountIs(1),   // $\ldots$ which are only given one argument $\ldots$
              // $\ldots$ which only argument (at index 0) is a deference
              hasArgument(0, Dereference)
            ));

        static const auto VarInitFromDereference =
          varDecl( // Match variable declarations $\ldots$
            anyOf( // $\ldots$ for which either of the following predicates are true $\ldots$
              hasInitializer(ignoringParenImpCasts(anyOf(
                // The variable has an initialiser expression, that is $\ldots$
                Dereference,                     // (see $\cref{tidy:initialisation/fundc}$)
                ConstructionExprWithDereference, // (see $\cref{tidy:initialisation/ctorc}$)
                initListExpr(hasDescendant(
                  // an initialiser list that has a sub-tree $\ldots$
                  Dereference))  // $\ldots$ that is a dereference (see $\cref{tidy:initialisation/list}$)
              ))),
              hasDescendant(expr(
                // Or there is a sub-tree which is a constructor call.
                ConstructionExprWithDereference // (see $\cref{tidy:initialisation/explicit}$)
              ))
            )).bind("InitedVar");
    \end{minted}
    \caption{The matcher expression that matches sub-trees for variable initialisations from the value of a pointer variable's dereference.}\label{tidy:init-matcher}
\end{listing}

\begin{listing}
    \begin{sublisting}[t]{.23\textwidth}
        \begin{minted}[bgcolor={}]{CPP}
            int i = p@$\rightarrow$@i;
        \end{minted}
        \caption{Fundamental ``C-style'' initialisation.}\label{tidy:initialisation/fundc}
    \end{sublisting}\hfill%
    \begin{sublisting}[t]{.23\textwidth}
        \begin{minted}[bgcolor={}]{CPP}
            T t = p@$\rightarrow$@t;
        \end{minted}
        \caption{Constructor call through ``C-style'' initialisation.}\label{tidy:initialisation/ctorc}
    \end{sublisting}\hfill%
    \begin{sublisting}[t]{.23\textwidth}
        \begin{minted}[bgcolor={}]{CPP}
            S s = {p@$\rightarrow$@i};
        \end{minted}
        \caption{List, direct, aggregate initialisation.}\label{tidy:initialisation/list}
    \end{sublisting}\hfill%
    \begin{sublisting}[t]{.23\textwidth}
        \begin{minted}[bgcolor={}]{CPP}
            T t(p@$\rightarrow$@t);
        \end{minted}
        \caption{Explicit constructor call expression.}\label{tidy:initialisation/explicit}
    \end{sublisting}
    \caption{The syntactically different forms of variable initialisation statements in \CC{}.}\label{tidy:initialisations}
\end{listing}

With the \linebreak matcher depicted in \cref{tidy:ptr-derefs}, we can now match dereferences, and we can move on to creating the matchers for the variable initialisations.
As with every syntactic element, the initialiser expression for a variable (\mintinline{CPP}{VarDecl}) is stored in the AST, and a special matcher, \mintinline{CPP}{hasInitializer()} exists to filter the details of the initialiser.
Unfortunately, just like dereferences, the initialisation of variables have multiple different methods and formats which mostly, but not entirely, correspond to different syntactic representations.
There are $11$ different initialisation schemes in \CC{}, but many of these differ only in their run-time semantics -- e.g.\ whether the memory behind the object is zero-filled before receiving the intended value.
As our goal here is to rewrite the initialisation in the source code, only the syntactically distinct forms of initialisation expressions are interesting to us, as depicted in \cref{tidy:initialisations}.
Whether a particular syntactic representation translates to one or the other semantics during code generation will be kept by the syntactic transformation.

We first define a helper match expression which detects every constructor call, which is from a dereference.
We will use this sub-expression in multiple cases of the full match expression.
The matcher corresponding to variable initialisations is presented in full in \cref{tidy:init-matcher}.
The full example contains some adaptor matchers (such as \mintinline{CPP}{ignoringParenImpCasts()}) which help ignore particular -- and from the perspective of this paper, unnecessary -- internal details of the \CC{} language.

\subsubsection{Matching guarded usage cases}
\begin{listing}[h]
    \begin{minted}[bgcolor={}]{CPP}
        static const auto FlowBreakingStmt = stmt(anyOf(
          // The built-in keyword-based statements and exception throw.
          returnStmt(), continueStmt(), breakStmt(), gotoStmt(),
          cxxThrowExpr(),

          // Function call where the called function is $[[\mathrm{noreturn}]]$.
          callExpr(callee(functionDecl(isNoReturn())))
        )).bind("EarlyReturn");

        static const auto Guard = ifStmt(
          // The condition of the branch $\ldots$
          hasCondition(allOf(
            // $\ldots$ contains a "usage" for a pointer variable $\ldots$
            hasDescendant(VarUsage.bind("UsedVar")),
            // $\ldots$ which is not a dereference!
            unless(hasDescendant(Dereference))
          )),

          // The content of the true branch is either $\ldots$
          hasThen(anyOf(
            FlowBreakingStmt, // a direct flow-away, e.g. if (p) return;
            compoundStmt(     // or a compound statement: if (p) { $\ldots$ }
              statementCountIs(1)  // With body of length 1 $\ldots$
              hasAnySubstatement(FlowBreakingStmt) // $\ldots$ that's a flow-away.
            )
          )),

          // And there is strictly no 'else' branch.
          unless(hasElse(stmt()))
        ).bind("GuardStmt");
    \end{minted}
    \caption{The match expression for guard statements on pointer variables.}\label{tidy:guard-stmt}
\end{listing}

By applying the matchers in the previous case, we can find the variable initialisations.
Now we can move on to implement the last set of matcher expressions that will handle guards.
We remind that a \emph{pointer variable guard} is a single simple conditional that checks the pointer's value and if the condition matches, breaks the execution of the function.
First, we create a helper matcher \mintinline{CPP}{FlowBreakingStmt} that matches such early breaks.
Then, and by reusing the matchers created in the previous sections, we can combine the matcher for the guards.
This case highlights how filtering matchers are used to narrow matches: an \mintinline{CPP}{ifStmt()} would match all \mintinline{CPP}{if}s, whereas by using \mintinline{CPP}{ifStmt(hasCondition(@$\ldots$@), hasThen(@$\ldots$@), hasElse(@$\ldots$@))} we can individually apply predicates to the three components of an \mintinline{CPP}{if}.
The full example is presented in \cref{tidy:guard-stmt}, with comments explaining the individual matches.

\subsection{Custom data structures and consuming the analysis results}\label{tidy:datastructure}
\begin{listing}
    \begin{minted}[bgcolor={}]{CPP}
        int* p = new int(); // $\leftarrow$ pointer variable
        foo(p);
        //  $\wedge$ usage point here
        return p@$\rightarrow$@x;
        //     $\wedge$ usage point here
        //     $\wedge\sim\sim$ dereference here
    \end{minted}
    \caption{Individual matchers match and fire their callback separately from each other.
        The code in this example totals $3$ \mintinline{CPP}{check()} callbacks.}\label{tidy:multiple-usages}
\end{listing}

When Clang-Tidy is running the analysis, every registered matcher expression will fire the \mintinline{CPP}{void check(const MatchFinder@$::$@MatchResult &Result);} \linebreak
    callback function.
The \mintinline{CPP}{Result} parameter contains the matched nodes in a string-to-node map, identified by the name they were \mintinline{CPP}{bind()} as in the match expression.
The conventional behaviour of the \mintinline{CPP}{check()} function is to retrieve the results from this map and emit a diagnostic with the \mintinline{CPP}{diag()} function.\footnote{%
    This function is available as every check inherits from \mintinline{CPP}{ClangTidyCheck}.
}

\begin{figure}
    \centering
    \includegraphics[height=.412\textheight, keepaspectratio]{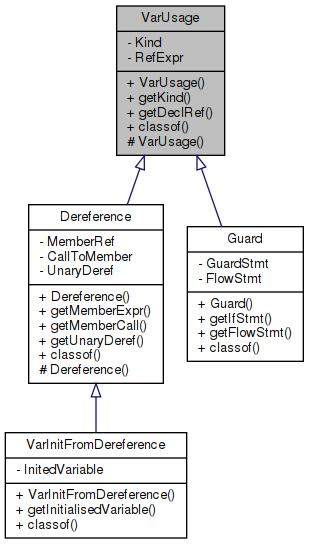}
    \caption{The UML class (and inheritance) diagram for the custom classes we implement to store usage contexts for pointer variables.
        These classes are only elaborated tuples, with no significant business logic embedded to them.}\label{tidy:ptr-use-classes}
\end{figure}

Our check's rules (see \cref{tidy}) have a typical pattern of \emph{``at most/only one point-of-use''}, which we cannot easily identify with matcher expressions that fire for every usage point.
Consider the example in \cref{tidy:multiple-usages}.
A callback is fired separately for both usages, and as such, we need to keep track of how many usages we identified for each variable.
Besides, due to the interwoven nature of our matcher expressions, the ``smaller'' matcher (e.g.\ \mintinline{CPP}{VarUsage}) fires the callback from the also matching ``larger'' matcher (e.g.\ \mintinline{CPP}{Dereference}), as depicted in \cref{tidy:matchers}.
This property of the analysis rule, however, allows us to emphasise that once the declarative and automated nature of performing the AST traversal and matching concluded, the control is ``back'' in the developer's hand, and can implement any logic to fine-tune the results further.
Clang-Tidy has seen an increase of such elaborate checks in recent years.

Thus, we will implement a counting machine in the \mintinline{CPP}{check()} function that will not immediately emit diagnostics, but instead, manage a data structure internal to the checker.
Thanks to the fact that the AST outlives the running checks, we can store just the pointers of the individual nodes in our ledger.
Naturally, the map's key will be the \mintinline{CPP}{VarDecl} itself.
For each pointer variable, we will store the list of usage contexts identified by the matchers in \cref{tidy:matcher-exprs}.
Because we are free to use every feature of \CC{} available to us, using inheritance and the object-oriented paradigm comes naturally -- the way the matchers build by using each other as sub-matchers, we also implement a type hierarchy (internal to the check's code) of usage contexts.
The schematic of the inheritance diagram can be seen in \cref{tidy:ptr-use-classes}.

The implementation will look as follows.
Each \mintinline{CPP}{check()} callback will analyse the matcher's results and construct a context object, based on what matcher matched.
This result is added to the usage list of the used variable.
In case a finer kind of matcher matches a usage point which was matched before by a broader one (such as in the second reference of \mintinline{CPP}{p} in \cref{tidy:multiple-usages}), the usage is transmogrified to the finer type.

\subsubsection{Inheritance hierarchies without run-time type information (RTTI) in LLVM}\label{tidy:casting}
Usually when it comes to inheritance and type-casting, \CC{} uses \mintinline{CPP}{dynamic_cast<U*>(p);} to cast a pointer of static type \mintinline{CPP}{T*} to \mintinline{CPP}{U*} if the run-time type is such.
However, the LLVM Compiler Infrastructure has adopted a method that does not require \emph{RTTI} to do casts.
This involves that instead of relying on the compiler to generate the necessary data structures, developers of class hierarchies are expected to implement it themselves.

\begin{listing}
\begin{minted}[bgcolor={},customlexer]{minted-llvm.py}
struct VarUsage {
  enum UsageKind { // The enum for LLVM-specific casts (see $\cref{tidy:casting}$)
    VU_Normal, VU_Dereference, VU_DerefInit, VU_Guard
  };

  // Public constructor for direct users.
  VarUsage(const DeclRefExpr* DRE) : VarUsage(DRE, VU_Normal) {}

  UsageKind getKind() const { return Kind; }
  const DeclRefExpr* getDeclRef() const { return RefExpr; }
  static bool classof(const VarUsage* VU) { return true; }
protected:
  // Delegated constructor for subclasses.
  VarUsage(const DeclRefExpr* DRE, UsageKind K) :
      Kind(K), RefExpr(DRE) {}
private:
  const UsageKind Kind;
  const DeclRefExpr* RefExpr;
};

struct Dereference : VarUsage { // A Dereference is-a Usage.
  Dereference(const DeclRefExpr* PtrVarReference, const Expr* DerefExpr)
    : Dereference(PtrVarReference, DerefExpr, VU_Dereference) {}

  static bool classof(const VarUsage* VU) {
    // Dereference and DerefInit are dereferences.
    return VU@$\rightarrow$@getKind() @$\ge$@ VU_Dereference &&
      VU@$\rightarrow$@getKind() @$\le$@ VU_DerefInit;
  }
  // $\ldots$ getters for the members $\ldots$
protected:
  Dereference(const DeclRefExpr* PtrVarReference,
    const Expr* DerefExpr, UsageKind K) : VarUsage(PtrVarReference, K),

      // Fill the members based on what dynamic type the
      // usage expression has.
      MemberRef(dyn_cast<MemberExpr>(DerefExpr)),
      CallToMember(dyn_cast<CXXMemberCallExpr>(DerefExpr)),
      UnaryDeref(dyn_cast<UnaryOperator>(DerefExpr)) {
      assert((MemberRef || CallToMember || UnaryDeref) &&
        "Invalid usage expression given, invalid dereference context!");
      if (UnaryDeref @$\ne$@ nullptr)
        assert(UnaryDeref@$\rightarrow$@getOpcode() @$==$@ UO_Star);
  }
private:
  const MemberExpr* MemberRef;
  const CXXMemberCallExpr* CallToMember;
  const UnaryOperator* UnaryDeref;
};
\end{minted}
\caption{Implementation of data members and representation using custom inheritance hierarchies, specific to LLVM flavour.}\label{tidy:custom-classes}
\end{listing}

This is done by creating an \mintinline{CPP}{enum} that contains an enum constant for all potential subclasses, and storing the value of this enum for all instances with a public \emph{getter}.
Additionally, developers are expected to implement the \mintinline[breaklines,customlexer]{minted-llvm.py}{static bool classof(const BaseClass*)} method for every derived class, which returns \mintinline{CPP}{true} if -- based on the value of this \emph{Kind} enum -- the current instance belongs to the subclass.
This way, LLVM's library function \mintinline[customlexer]{minted-llvm.py}{dyn_cast<U>(p)} can be used to achieve the same result as \mintinline{CPP}{dynamic_cast} does in everyday \CC{} programs.

\subsubsection{Extending the check class \mintinline{CPP}{RedundantPointerCheck}}
Before we implement the callback mechanisms, we need to extend the \linebreak
    \texttt{readability/RedundantPointerChecker.h} file with the new classes and add the data structure to the check class's body.
We create $4$ classes: \mintinline{CPP}{VarUsage}, \mintinline{CPP}{Dereference}, \mintinline{CPP}{VarInitFromDereference}, which inherit from each other in this order, and \mintinline{CPP}{Guard} that inherits from \mintinline{CPP}{VarUsage}.
The inheritances correspond to how the matchers embed each other as sub-matchers.
A piece of the full implementation, with the necessary methods elaborated, is in \cref{tidy:custom-classes}.

To add the instances of our model to the check's run-time, in LLVM projects, it is customary to use LLVM's data structures found in the \texttt{llvm/ADT} library.
We will use \mintinline{CPP}{llvm@$::$@DenseMap<K, V>} and \mintinline{CPP}{llvm@$::$@SmallVector<T, N>} for this implementation.
The former is a hash-table based key-value map which works well for easily hashable keys (such as pointers, in our case).
The latter is a locality optimised vector, where at most $N$ elements are allocated inside the vector, not to a buffer on the heap.
The full type of a new data member that should be added to \mintinline{CPP}{RedundantPointerCheck} is \mintinline[breaklines]{CPP}{llvm@$::$@DenseMap<const VarDecl*, @\\@llvm@$::$@SmallVector<VarUsage*, 4>}.\footnote{%
    $4$ is an arbitrary number here, but a good approximation to achieve reasonable performance with the check.
}
These classes mostly follow the usual API of \texttt{<map>} and \texttt{<vector>} from the \CC{} Standard Template Library.

As seen in \cref{tidy:custom-classes}, each usage context class takes the \mintinline{CPP}{DeclRefExpr} -- the pointer variable's reference in an expression -- as a constructor parameter.
The \mintinline{CPP}{check()} method will be responsible for constructing these classes; however, for clarity, we suggest adding a private member function to \mintinline{CPP}{RedundantPointerCheck}: \mintinline{CPP}{void addUsage(const DeclRefExpr* DRE, VarUsage* Usage);}.
This \linebreak
    method will add the given usage to the vector of usages for the referenced variable (which can be obtained via a call and a cast: \mintinline[customlexer]{minted-llvm.py}{cast<VarDecl>(DRE@$\rightarrow$@getDecl())}), or in case the same \mintinline{CPP}{DRE} is already added, and the new usage is a \emph{more specialised one}\footnote{%
    Such as in the case \cref{tidy:multiple-usages}, where for the second usage, \mintinline{CPP}{addUsage()} is called \emph{twice}.
} the usage is replaced.
The implementation for this logic is trivial, and the LLVM specifics follow from previous sections, and as such, we omit elaborating it here.

\subsubsection{Implementing the \mintinline{CPP}{check()} callback}
\begin{listing}
    \begin{minted}[bgcolor={}]{CPP}
        // Handle the matcher from $\cref{tidy:guard-stmt}\ldots$
        if (const auto* Guard = Result.Nodes.getNodeAs<IfStmt>("GuardStmt")) {
          const auto* Flow = Result.Nodes.getNodeAs<Stmt>("EarlyReturn");
          const auto* DRE  = Result.Nodes.getNodeAs<DeclRefExpr>("UsedVar");

          addUsage(DRE, new Guard(DRE, Guard, Flow));
          return;
        }
        // Handle the matcher from $\cref{tidy:init-matcher}\ldots$
        if (const auto* VarInit =
            Result.Nodes.getNodeAs<VarDecl>("InitedVar")) {
          const auto* DerefExpr = Result.Nodes.getNodeAs<Expr>("DerefUsage");
          const auto* DRE = Result.Nodes.getNodeAs<DeclRefExpr>("DerefdVar");

          addUsage(DRE, new VarInitFromDereference(DRE, DerefExpr, VarInit));
          return;
        }
        // $\vdots$
    \end{minted}
    \caption{Using early \mintinline{CPP}{return}s and sub-to-base order of result handling for the \mintinline{CPP}{check()} function.}\label{tidy:check()}
\end{listing}

Given the steps taken in previous sections to encode the contextual information in custom data structures and handling logic, the \mintinline{CPP}{check()} function's implementation to fill the data structures of our model becomes easy.
Note that the function is fired for every top-level matcher (matchers that were given to \mintinline{CPP}{addMatcher()}, see \cref{tidy:matching}), irrespective of how a top-level matcher might match the sub-expression matched by another top-level matcher.
Due to this, care must be taken to fetch the more specialised results first, every time the function is called.
This is achieved by organising the code into a series of \mintinline[breaklines]{CPP}{if (const auto* X = @$\linebreak$@ Result.Nodes.getNodeAs<T>(@$\beta$@))} conditionals, all which \mintinline{CPP}{return} at the end of the true branch.
$\beta$ refers to the string identifier that was given to the \mintinline{CPP}{bind()} method of the matcher, and \texttt{T} is the node type that is matched.
The semantics of \mintinline{CPP}{getNodeAs()} specify that it will return a \mintinline{CPP}{T*} if and only if a node named $\beta$ is matched, and it can be converted to $T$,\footnote{%
    Conversion is done via \mintinline[customlexer]{minted-llvm.py}{dyn_cast<T>} under the hood, as all class hierarchies in LLVM implement a (more complex form of) \mintinline[customlexer]{minted-llvm.py}{classof()} function, as discussed in \cref{tidy:casting}.
} otherwise \mintinline{CPP}{nullptr} is returned, and the branch is not taken.
This is similar to how \mintinline{CPP}{catch} ``branches'' are organised in case different levels of an exception hierarchy is to be handled -- the more specialised subclass has to be attempted first.

An example excerpt from the implementation can be seen in \cref{tidy:check()}.
The remaining cases in the method can be implemented analogously.
Typically, the responsibility of emitting the diagnostics and offering automated fixes are also the responsibility of \mintinline{CPP}{check()}.
However, in our current implementation, we had to defer this to a later stage in the analysis due to our use of custom modelling.

\begin{algorithm}
    \DontPrintSemicolon
    \KwIn{usages of pointer variable \texttt{p} modelled by \mintinline{CPP}{check()}}

    \SetKwFunction{Code}{initSourceCode}
    \SetKwFunction{DefCtor}{is\_default\_constructible}
    \SetKwFunction{Assign}{is\_assignable}
    \SetKwFunction{Var}{Var}
    \SetKwFunction{Diag}{Diagnose}
    \SetKwFunction{Fix}{EmitFix}
    \SetKwFunction{Usage}{usage}
    \SetKwFunction{GuardC}{condition}
    \SetKwFunction{GuardF}{flow}

    \uIf{$\#usages(\texttt{p}) = 0 \vee \#usages(\texttt{p}) \ge 3$}{%
        \KwRet{Nothing}\;
    }

    $\mathcal{C} \gets$ \Code{p}

    \If(\tcp*[h]{exactly one usage: replace with p's value}){$\#usages(\texttt{p}) = 1$}{%
        \Diag{p, "Redundant pointer variable with only one usage"}\;
        \Fix{p, $\emptyset$}             \tcp*{delete p's declaration}
        \Diag{\Usage{p}, "Pointer usage location"}\;
        \Fix{\Usage{p}, $\mathcal{C}$}   \tcp*{put p's value to the usage point}
    }

    \If{$\#usages(\texttt{p}) = 2 \wedge \left(\exists! g <:\right.$\mintinline{CPP}{Guard}$\left.\right) \wedge$ $\left(\exists u <:\right.$\mintinline{CPP}{VarInitFromDereference}$\left.\right)$ $\wedge$ $\mathrm{\CC{}17}$ $\wedge$ \DefCtor{\Var{u}} $\wedge$ \Assign{\Var{u}}}{%
            \Diag{p, "Redundant pointer variable declared"}\;
            \Diag{\Usage{p}, "Variable dereferenced here, swap variables"}\;
            \Fix{p, \Var{u}}           \tcp*[r]{rewrite pointer var to initialised var}
            \;
            \Diag{u, "after swap, the initialisation is not needed at this location"}\;
            \Fix{u, $\emptyset$}       \tcp*[r]{the line of var init is not needed}
            \;
            \Diag{g, "rewrite the conditional to \CC{}17 initialise the pointer"}\;
            $\Pi \gets$ "(!\GuardC{g} \mintinline{CPP}{||} ((u, \mintinline{CPP}{false})))"\;
            \Fix{g, "if (\texttt{p} = $\mathcal{C}$; $\Pi$)"}   \tcp*[r]{rewrite the condition to scoped init}
    }

    \caption{The algorithm of the decision points the diagnostic-emitting part of the check takes.}\label{tidy:checker-algo}
\end{algorithm}

\subsection{Emitting diagnostics and creating automatic fixes}
In the current state of our check, we do not have any diagnosis-emitting callbacks firing, as we reserved the \mintinline{CPP}{check()} function for the generation of the usage model.
However \mintinline{CPP}{ClangTidyCheck}, the class every check implementation derives from, offers a few more overrideable functions, such as \mintinline[breaklines,breakafter=d]{CPP}{void onEndOfTranslationUnit();}\footnote{%
    Similarly, a \mintinline{CPP}{void onStartOfTranslationUnit();} exists, although it's usefulness for Tidy checks is not yet exploited, as each Clang-Tidy invocation runs on \textbf{one} translation unit, like the compiler.
} which we will use to consume the previously generated model.
As the name suggests, the implementation in this function will fire once for each translation unit handled, after the translation unit has been processed by the tool -- in Tidy's terms, this means that all matching and \mintinline{CPP}{check()} calls are done.

Per the targeted use cases for the check, as discussed in \cref{tidy}, we will take the following steps for generating an output.
The algorithmic skeleton is outlined in \cref{tidy:checker-algo}.
In the first case, for variables with exactly one usage, we emit the code of the pointer's initialisation to the only usage point and remove the pointer variable from the code.
In the second case, if applicable, we will perform the swapping of variables and initialisation as detailed in \cref{tidy:deref-init-rewrite}.

\begin{listing}
    \begin{minted}[bgcolor={}]{CPP}
        Lexer@$::$@getSourceText(
            CharSourceRange@$::$@getCharRange(@$B$@,
                Lexer@$::$@getLocForEndOfToken(@$E$@, 0, SM, LO),
            ), SM, LO);
    \end{minted}
    \caption{The implementation to get the source code between two \mintinline{CPP}{SourceLocation}s, $B$ and $E$.}\label{tidy:full-source-text}
\end{listing}

In the following, we detail the steps on how to implement the algorithm.
To interface with the language context and source code itself, the \mintinline{CPP}{ASTContext} needs to be obtained.
This is retrieved from every \mintinline{CPP}{Decl} via the \mintinline{CPP}{getASTContext()} method.
This context object stores a reference to the 
    \mintinline{CPP}{SourceManager} instance through \mintinline{CPP}{getSourceManager()}, through which the Clang libraries can access the translation unit's textual buffer.
In addition, the standard's version and the configuration of language features and potential extensions are also obtainable from the context, via \mintinline{CPP}{getLangOpts()}.
The \mintinline{CPP}{LangOptions} instance is a simple collection of \mintinline{CPP}{bool} fields, toggled if the respective feature is enabled.
In the following, we will refer to these context objects as \mintinline{CPP}{Ctx}, \mintinline{CPP}{SM}, and \mintinline{CPP}{LO}, respectively.
For example, whether the current translation unit is compiled with the \CC{}17 standard is queried by checking \mintinline{CPP}{LO.CPlusPlus17}'s truth value.

\subsubsection{Fetching the initialising expression for a variable}
Unfortunately, we have to provide the replacement text as a string.
However, this replacement in our case comes from another location in the analysed code, and as such, we must manually fetch it.
We will achieve this through the \mintinline{CPP}{Lexer} class's methods, which usually require a \mintinline{CPP}{SourceLocation} or \mintinline{CPP}{SourceRange} object as their argument.
A \mintinline{CPP}{SourceLocation} is the identifying entity for each token or character's location in the translation unit's input text, and querying and manipulation through these objects are commonplace for tools that manipulate the source code itself.
A \mintinline{CPP}{SourceRange} is a pair of locations that highlight a range.
Almost all nodes in the AST are eventually assigned with at least two, but usually more \mintinline{CPP}{SourceLocation}s, namely there are at least two query functions, \mintinline{CPP}{getBeginLoc()} and \mintinline{CPP}{getEndLoc()}.\footnote{%
    Various AST node types store and allow getting of additional locations, e.g.\ in the case of \mintinline{CPP}{BinaryOperator}, the exact location of the operator token as written in the code can be obtained via \mintinline{CPP}{getOperatorLoc()} which will be a location between \emph{begin} and \emph{end}.
}

Unfortunately, the location for the \emph{end} of a node, when converted to a character in the source buffer often ends up as the \textbf{first} character of the token that formed the last token that comprises the node.
This is a perhaps unsavoury detail that results from the fact that lexical analysis and parsing of modern programming languages (see \cref{handout:lexical-analysis}) is a complex task.
The \textbf{full} character range between two \mintinline{CPP}{SourceLocation}s, $B$ and $E$ can be obtained with the code snippet that is seen in \cref{tidy:full-source-text}.

\begin{listing}
    \begin{minted}[bgcolor={}]{CPP}
        VarDecl *Var = @$\ldots$@; // points to a variable in the AST, e.g. match result
        diag(Var@$\rightarrow$@getLocation(), "variable: %0", DiagnosticIDs@$::$@Warning) @$<\!\!<$@ Var;
    \end{minted}
    \begin{minted}[bgcolor={}]{Text}
        example.cpp:1:5: @\color{royalpurple}{warning:}@ variable: 'Var' @\color{darkgray}{\emph{[my-tidy-check]}}@
         int Var;
             @$\wedge$@~~
    \end{minted}
    \caption{The usage of the \mintinline{CPP}{diag()} function to produce a dummy diagnostic to a variable, and the output, as seen on the user's standard error stream, from executing the call.}\label{tidy:diag()}
\end{listing}

\subsubsection{Creating diagnostics with \mintinline{CPP}{diag()}}
All \mintinline{CPP}{ClangTidyCheck} subclasses inherit the \mintinline{CPP}{diag()} method, which is used to emit diagnostics through the standard error stream to the user, or specialised tools, such as CodeChecker (see \cref{codechecker}).
These diagnostics are presented and behave the same way as the standard compiler diagnostics (see \cref{handout:compiler-errors}) that the users might get from an ill-formed translation unit.
The \mintinline{CPP}{diag()} function takes three arguments, in order: first, the \mintinline{CPP}{SourceLocation} where the source line and column numbers, and the caret (\^{}) of the diagnostic itself should be placed, the diagnostic message, and the diagnostic's severity.
In Clang-Tidy, it is customary to use the \mintinline{CPP}{DiagnosticIDs@$::$@Warning} and \mintinline{CPP}{DiagnosticIDs@$::$@Note} severities.
The diagnostic's message might contain placeholders (\texttt{\%0}, \texttt{\%1}, \ldots)\footnote{%
    More elaborate placeholders, such as automatic selection based on a numeric value, automatically formatted ordinals (\emph{1st}, \emph{2nd}, \ldots) can be encoded in the diagnostic's message ``template'' which is expanded by Clang when the diagnostic is printed.
    Please refer to the \emph{``Clang CFE Internals Manual''} document for the exact list of format strings in case your check's elaborate enough to require such.
    Most Clang-Tidy checks only emit the matched symbol's (such as a function's) name.
} to which symbol names can be inserted based on the name of the symbol that matched.

Once the \mintinline{CPP}{diag()} function is called, the tool is set into diagnostic generation mode.
The diagnostics will be printed to the user at the end of the full expression where the object returned by the function call is destructed.
It is important that no two diagnostics are ``in flight`` at the same time.
The returned temporary object can be fed with diagnostic pieces via \mintinline{CPP}{operator @$<\!\!<$@}, similarly to how \mintinline{CPP}{std@$::$@cout} is written to.
The placeholders mentioned above are filled in the order of these feed operations.
In case an AST node's pointer is given to the operator, the name of the AST node (if applicable) is automatically written in, pretty-printed.
An example invocation of the \mintinline{CPP}{diag()} function can be seen in \cref{tidy:diag()}.

The \texttt{Diagnose} lines in \cref{tidy:checker-algo} correspond to \mintinline{CPP}{diag()} invocations.

\subsubsection{Creating automated fixes via \mintinline{CPP}{FixItHint}}
The biggest highlight of syntax-tree-based analysis and Clang-Tidy comes from the ability to offer rewrites, in Clang terms, \emph{FixIt}s, which are presented to the user together with the diagnostic.
The tool can also automatically apply it to the source code if the user instructs it to do so.
An arbitrary number of \emph{FixIt}s can be applied to each diagnostic output line, as depicted in \cref{tidy:fixit}.
However, it is customary to use only one per diagnostic, at least for Clang-Tidy checks.
A single FixIt can be one of three kinds: insertion, replacement, or removal.
FixIts are instantiated through the factory functions \mintinline{CPP}{FixItHint@$::$@CreateInsertion(Loc, Text)}, \mintinline{CPP}{FixItHint@$::$@CreateReplacement(Range, NewText)}, \linebreak
    and \mintinline{CPP}{FixItHint@$::$@CreateRemoval(Range)}, respectively.

\begin{figure}
    \centering
    \includegraphics[width=\textwidth]{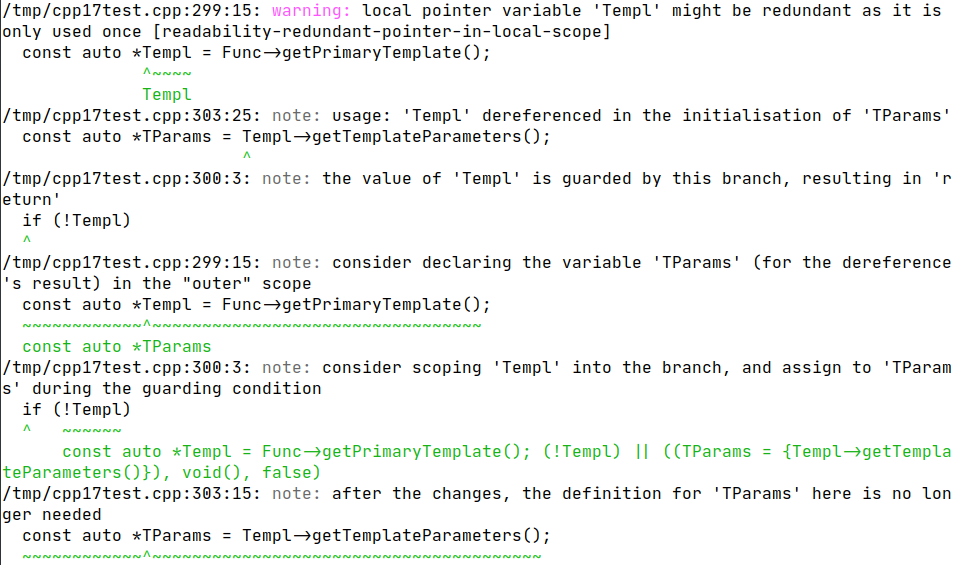}
    \caption{The output of the \emph{redundant pointer variable} check for \CC{}17 guarded initialisation rewrites, complete with the showing of individual matches, and \emph{FixIt}s.}\label{tidy:output}
\end{figure}

If Clang-Tidy is invoked with the \texttt{-fix} flag, the suggested rewrites are automatically applied after analysis to the input translation unit's code.
Most checks aimed at semantically correct refactoring, such as the \texttt{modernize} check group, offer such automated fixes.
The \texttt{EmitFix} lines in \cref{tidy:checker-algo} refer to the generation of these \mintinline{CPP}{FixItHint}s.
An excerpt from the implementation that rewrites the \emph{guard branch}'s condition is shown in \cref{tidy:guard-rewrite}.
The complete output of the check for an example test code can be seen in \cref{tidy:output}.

\begin{listing}
    \begin{minted}[bgcolor={}]{CPP}
        VarDecl* Var = @$\ldots$@; // points to a variable in the AST
        diag(Var@$\rightarrow$@getLocation(), "variable: %0 should be prefixed with 'my'",
            DiagnosticIDs@$::$@Warning) @$<\!\!<$@ Var
            @$<\!\!<$@ FixItHint@$::$@CreateInsertion(Var@$\rightarrow$@getLocation(), "my");
    \end{minted}
    \begin{minted}[bgcolor={}]{Text}
        example.cpp:1:5: @\color{royalpurple}{warning:}@ variable: 'Var' should be prefixed with 'my'
         int Var;
             @$\wedge$@~~
             @\color{darkspringgreen}{my}@
    \end{minted}
    \caption{The example in \cref{tidy:diag()} extended with automated fix offering.
        Through feeding \mintinline{CPP}{FixItHint}s, Tidy can print source code change suggestions to the user.}\label{tidy:fixit}
\end{listing}

\begin{listing}
\begin{minted}[bgcolor={}]{CPP}
// Suppose these two instances of our custom model contains the
// necessary AST references. (See $\cref{tidy:datastructure}$)
const VarInitFromDereference* Init = @$\ldots$@;
const Guard* GuardBranch = @$\ldots$@;

// The initialisation code for the original pointer variable:
//     T* p = init_ptr();
std@$::$@string PtrInit = sourceCode(Init@$\rightarrow$@getDeclRef()@$\rightarrow$@getDecl());

// The initialisation code for the variable created from the dereference:
//     = p$\rightarrow$x;
std@$::$@string VarInit = initSourceCodeForVar(Init@$\rightarrow$@getInitialisedVariable());

llvm@$::$@SmallString<384> NewCondition = "; ("; // Start the condition.

// Append "!$\phi$(p)" (see $\cref{tidy:deref-init-rewrite}$)
NewCondition.append("!(");
NewCondition.append(sourceCode(GuardBranch@$\rightarrow$@getIfStmt()@$\rightarrow$@getCond()));
NewCondition.append(")");

// Append "|| (assignment)"
NewCondition.append(" || ((");
NewCondition.append(Init@$\rightarrow$@getInitialisedVariable()@$\rightarrow$@getName());
NewCondition.append(VarInit);
NewCondition.append("), false)";

NewCondition.append(")"); // Close the condition.

// Append the full condition after the pointer's declaration code.
PtrInit.append(NewCondition.c_str());

// Now PtrInit's contents are:
//     T* p = init_ptr(); (!(p == nullptr) || ((v = p$\rightarrow$x), false))

diag(Guard@$\rightarrow$@getIfStmt()@$\rightarrow$@getIfLoc(),
    "rewrite if() to initialise the pointer in the branch scope",
    DiagnosticIDs@$::$@Warning)
    @$<\!\!<$@ FixItHint@$::$@CreateReplacement(
        // We only rewrite the $\textbf{condition}$, so the if's outer () remain intact.
        Guard@$\rightarrow$@getIfStmt()@$\rightarrow$@getCond()@$\rightarrow$@getSourceRange(),
        PtrInit);
\end{minted}
\caption{The LLVM/Clang object model and internal API offers a comfortable and concise way to implement even complex source code transformations, such as rewriting pointer dereference guards.
    \mintinline{CPP}{llvm@$::$@SmallString<@$N$@>} is another custom LLVM data structure, which offers most of the usual \mintinline{CPP}{std@$::$@string} API.
    However, it allocates the first $N$ characters in place, as opposed to putting them into a buffer on the heap, allowing for, in our case, quicker concatenations.}\label{tidy:guard-rewrite}
\end{listing}

\clearpage
\subsection{Exercises for the Reader}
In the previous sections, we have shown how to use syntax-tree based match expressions to find C and \CC{} code parts and refactor, modernise or diagnose them based on the implemented criteria.
The example of \emph{redundant pointer variables} was complex enough to allow us to detail the ``nooks and crannies'' of the AST Matcher library and Clang-Tidy as a whole, it is far from complete.
There are some cases which we have implemented, but omitted discussing in this paper, and may be viewed as \emph{future work}, were someone to implement only the contents of this section.
These additional exercises allow for a deeper dive into the details of the AST and the \CC{} language itself.
As firm believers in the notion that industrially viable software can only be created via active work, we encourage the Reader to try these exercises, always keeping the documentation for matchers, the AST node types and their methods, and a set of test examples handy while developing.
We indicate with $\star$ the ``exercises'' that are ``harder'', but extremely rewarding, and might require significant thought and refactoring of the currently implemented version.

\begin{figure}[H]
    \centering
    \begin{turn}{-90} 
        \begin{minipage}{.995\textwidth}
            \includegraphics[height=.6\textwidth,keepaspectratio]{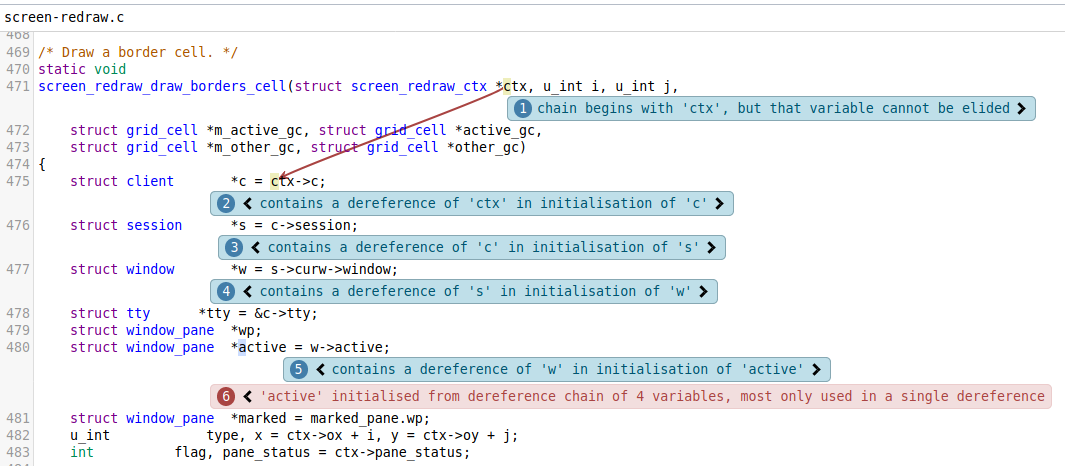}
        \vspace{.2cm}
        \captionof{figure}{The match result for variable-initialising pointer dereference \textbf{chains} on the \emph{TMux} project, as visualised by CodeChecker (see \cref{codechecker}).
            Image from~\cite{Szalay2020Towards}.}\label{tidy:tmux-chain}
        \end{minipage}
    \end{turn}
\end{figure}

\begin{itemize}
    \item Note that initially we only considered variables to be \emph{pointer variables} if they were of a pointer type.
        This matcher does not apply to constructs such as \mintinline{CPP}{auto p = new int();}, as this variable is of \mintinline{CPP}{auto}-type, which, after the semantic analysis also happens to be a pointer, but this is an additional indirection in the type system.
    \item \CC{} allows the users to define custom types, which also may work akin to, or in place of, pointers.
        \emph{Iterators} are the most famous example of this.~\cite{BrunnerICAI2017}
        However, these types, which we named \emph{``dereferenceables''} are also not matched by the \mintinline{CPP}{hasType(pointerType())} matcher, and as such, need to be found and analysed separately.
        (Hint: it is enough to consider the existence of the unary $\star$ and the $\rightarrow$ operator for such types.)
    \item Naturally, \mintinline{CPP}{auto} variables of \emph{dereferenceable type} should also be handled.
    \item In some cases, it is ill-advised to diagnose or destructively change the code by rewriting.
        One such case is library headers that are outside of the project's (or any users') control, such as \emph{system headers} and the STL.
    \item $\star$ Certain \CC{} constructs, such as the ``range-based \mintinline{CPP}{for} loop'' (\mintinline{CSharp}{foreach}) introduce pointer or dereferenceable local variables that are syntactic sugar and do not exist in the source code, only created by the parser.
        These should be ignored, together with any loop variable that also happens to be a pointer, as no sensible rewrite exists for these.
        (Hint: the \mintinline{CPP}{hasParent()} matcher can be used to filter matches by an upwards traversal in the AST, but this act is computationally intensive.)
    \item The previous statement about the inability to succinctly rewrite also applies to function parameters (\mintinline{CPP}{ParmVarDecl}s).
    \item Usages that match inside \emph{macro expressions} are also not rewritable properly and should be ignored.
        For this, Clang's \mintinline{CPP}{Preprocessor} layer can be used.
    \item $\star$ Consider that ultimately sometimes multiple pointers are used only once, and individual rewrite of each to one less variable is a costly operation, even for automated tools, as such ``chains'' are witnesses of deeper lying API design defects.
        The current implementation can be extended with additional modelling to identifier \emph{pointer dereference chains} that might be elided in one step.
        The details on this can be found in~\cite{Szalay2020Towards}, and one example result is depicted in \cref{tidy:tmux-chain}.
\end{itemize}

\vspace{2em}

\clearpage
\section{Symbolic Execution}\label{sa}

In the previous \cref{tidy} we saw the applicability of AST-based syntactic analysis.
In this section, we show some of its shortcomings, and describe an alternative technique called symbolic execution.

\subsection{An example: dangling pointers}\label{sa:example}

Memory errors are the culprits behind 70\% of the security vulnerabilities fixed year by year at information technology megacorporations like Microsoft~\cite{Microsoft} and Google~\cite{Chrome}.
In this chapter, let us help them (and the wider C/\CC{} community) by writing a static analysis module that finds a common type of memory error: dangling pointers.

The traditional, C-style version of this problem can be seen in \cref{sa:use-after-free-c}.
Here, we allocate a block of memory, referred to by the pointer variable \mintinline{CPP}{s}, then copy \mintinline{CPP}{s} to a new pointer called \mintinline{CPP}{c}.
This copy is not necessary, but it allows us to compare this example to the next, \CC{}-style example (\cref{sa:use-after-free-cpp}) line by line.
After the memory block is freed (in other words, \emph{released}), \mintinline{CPP}{s} and \mintinline{CPP}{c} both become \emph{dangling pointers}.
Any use of these pointers is considered a serious memory error.

\begin{listing}[h]
    \begin{minted}[linenos]{CPP}
        #include <cstdlib>

        char *useAfterFree() {
          char *s = (char *) std::malloc(10 * sizeof(char));
                        // Note: Memory is allocated
          char *c = s;  @$\label{lin:dangling}$@
          std::free(s); // Note: Memory is released
          return c;     // Warning: Use of memory after it is freed
        }
    \end{minted}
    \caption{A C-style use-after-free error.}\label{sa:use-after-free-c}
\end{listing}

Fortunately for the users, this \mintinline{CPP}{malloc}/\mintinline{CPP}{free} version of the dangling pointer problem has been around for such a long time that most bug-finding tools have already developed rules to find it.
In the comments of \cref{sa:use-after-free-c}, you can see a textual representation of the bug report given by a popular analysis tool that will be introduced in \cref{sa:csa}.

Fortunately for static analysis researchers, however, a new rendition of the problem always arises in new standards of \CC{}.

The \mintinline{CPP}{std@$::$@string} container\footnote{%
    \mintinline{CPP}{std@$::$@string} is a specialization of the \mintinline{CPP}{std@$::$@basic_string} class template for \mintinline{CPP}{char}-type sequences.
    Other variations exist for other character types, such as \mintinline{CPP}{wchar_t}, \mintinline{CPP}{char32_t}, etc.%
} of the \emph{\CC{} Standard Template Library} (STL) exhibits a similar kind of vulnerability to \mintinline{CPP}{malloc}/\mintinline{CPP}{free}.
Its \mintinline{CPP}{c_str} and \mintinline{CPP}{data} methods allow the user to obtain a raw pointer pointing to the inner buffer of the container that holds the actual character sequence.

In our example in \cref{sa:use-after-free-cpp}, this pointer is called \mintinline{CPP}{c}.
The \CC{} standard specifies the conditions under which all pointers referring to the elements of a \mintinline{CPP}{std@$::$@string} sequence become \emph{invalidated}~\cite{StdDraft}, and calling a method like \mintinline{CPP}{clear} on the string is one of them.
(STL implementations are free  to reallocate the string at this point.)
This means that after Line~\ref{lin:dangling}, \mintinline{CPP}{c} is a dangling pointer, and its subsequent use is a programming error.

\begin{listing}[h]
    \begin{minted}{CPP}
        #include <string>

        const char *useAfterClear() {
          std::string s = "hello";
          const char *c = s.c_str();
          s.clear();    // c is invalidated.
          return c;     // We should report this error.
        }
    \end{minted}
    \caption{A \CC{}-style \emph{use-after-free} error.}\label{sa:use-after-free-cpp}
\end{listing}

Let us think about a static analysis check to find this error by using the AST-based analysis we learned in the previous chapter.
One common case we might consider is the following:

\vspace{5pt}
\mintinline{CPP}{return std::to_string(name).c_str();}
\vspace{5pt}

\noindent We might deduce that using the buffer pointer of a temporary \mintinline{CPP}{std@$::$@string} is an error, and try to express it with AST matchers.
However, we will find that some cases that fit this description are actually correct:

\vspace{5pt}
\mintinline{CPP}{return strcpy(dest, std::to_string(name).c_str());}
\vspace{5pt}

\noindent In this example, the buffer pointer of a temporary string is passed to a function call as an argument.
This is a valid use of the pointer, because the string will only be destroyed at the end of the full-expression (i.e.\ after the \mintinline{CPP}{strcpy()} call has been evaluated), just like the pointer itself.
We would have to exclude this counter-example in our AST matchers to avoid giving an incorrect report to the user.

Avoiding such bogus reports would make our checker increasingly complicated.
However, we should also worry about losing useful reports.
Consider the example in \cref{sa:path-sensitive-bug}.

\begin{listing}[h!]
    \begin{minted}{CPP}
        const char *hardToMatchInAST(bool cond) {
          std@$::$@string s;
          if (cond)
            return s.c_str(); // The error is caused here, $\ldots$
          // $\ldots$
        } // $\ldots$ but it actually manifests here.
    \end{minted}
    \caption{A possible false negative case of AST-based analysis.
        Although the return statement is in the beginning of the function, the error occurs at the end of the function, when the string is destroyed.
        The unknown amount of code between them makes this case difficult to find with syntactic analysis.}\label{sa:path-sensitive-bug}
\end{listing}

In this case, the buffer pointer is returned quite early in the function, but it will only become dangling at the end of it.
Although these two events happen tightly after each other during a real execution, they are arbitrarily far from each other in the AST of the function.

The compiler's goal with the AST is to represent the source code as practically and faithfully as possible, with additional semantic and diagnostic annotations.
It is not the goal of the compiler to simulate what is actually happening in the program.
Syntactic bug-finding tools look for patterns in the AST, but cannot reason about the values of the variables, or about which paths a real execution may take through the program.
For these, we need a more powerful analysis method.

\subsection{Symbolic execution}

\emph{Symbolic execution} was first described by IBM researcher James King~\cite{king1976symbolic}.
It is a static program analysis technique that works by inspecting the source code, just like AST matching (see \cref{tidy}).
However, symbolic execution tries to understand what would be happening in the program if it was run.
It interprets the program using \emph{symbolic execution semantics} - usual execution semantics extended to work for arbitrary \emph{symbols} that may substitute for real data values in the program.
Such a substitution is needed whenever we encounter an unknown input value during the simulation process.
We assume that the values of variables can be expressed as a function of these symbols at any point in the program.

An important characteristic of symbolic execution is \emph{path-sensitivity}.
At each conditional statement, the analysis engine will attempt to decide whether a given branch can ever be taken during a real execution run, and will abandon branches that are known to be infeasible.
However, conditions may contain symbols.
In that case, the analyser records the symbolic constraints required to take a particular branch in the so-called \emph{path condition}.
Branches with different sets of constraints are pursued further separately, giving us an exponential algorithm.

In the remaining sections, we will use a symbolic execution framework called the \emph{Clang Static Analyzer} to create our own bug-finding tool for the dangling pointer problem.

\subsubsection{The Clang Static Analyzer}\label{sa:csa}

Based on our description of symbolic execution earlier, it is hard to imagine how the method works in realistic settings.
To aid our imagination, in this section we introduce a symbolic execution toolset that is used by companies like Apple~\cite{AppleAnalyze} and Google~\cite{ChromeAnalyze} every day.

The Clang Static Analyzer~\cite{kremenek2008finding} (we will abbreviate it as \emph{the Clang Analyzer}) is an open-source analysis framework built into the popular C/\CC{}/Objective-C compiler, Clang~\cite{lattner2008llvm}.
Clang is part of the wider LLVM Compiler Infrastructure Project~\cite{lattner2004llvm}.
Its static analysis toolset is not only built into IDEs like Apple's XCode and Microsoft's  Visual Studio, but its source code can also be downloaded from the official repository (for instructions, see~\cite{ClangDoc}).

The Clang Analyzer records the progression of its analysis in a data structure called the \emph{exploded graph}.
Each vertex of this graph represents the \emph{symbolic program state} at a given \emph{program point}.
The program point determines the current location in the program, similarly to an instruction pointer.
The symbolic program state corresponds to a set of real program states.
The edges of the graph are transitions between the states.
Memory is represented using a hierarchy of memory regions described in~\cite{xu2010memory}.

A symbolic state consists of the following components~\cite{dergachev2016clang}:
\begin{itemize}
\setlength\itemsep{0pt}
    \item the \textbf{environment}, which is a mapping from source code expressions to symbolic expressions,
    \item the \textbf{store}, which maps memory locations to symbolic expressions,
    \item \textbf{range constraints}, showing the known value ranges of symbols, and
    \item a \textbf{generic data map} that stores checker-specific information.
\end{itemize}

\noindent The program state can also carry \textbf{taint} information on symbols that come from insecure sources, used by the \mintinline{CPP}{GenericTaintChecker} module.

\emph{Checkers} are plug-in modules that cooperate with the engine during the analysis process.
They can subscribe to various events in the program, choosing to run before or after them.
They may query and modify the program state, storing extra data in it as needed.
Their main purpose is to define the conditions that imply the presence of an error in the analysed program.
If a given program point matches those conditions, they create a bug report.
If the issue is critical, they also terminate the analysis on that execution path.
The bug report includes a representation of the execution path that leads to the error, aiding users in understanding the problem.
Reports are collected and processed after the analysis is finished on each path.

The question of where to start the analysis is already not trivial.
The \mintinline{CPP}{main} function might not be available, or in case of libraries, might not exist at all.
The Analyzer therefore picks a function and starts to interpret it without a calling context.
Such a function is called a \emph{top-level function}.
When it encounters a call to a function with a known body, the analyser can continue the analysis inside the callee, preserving all the information known at the call site.
This is called \emph{inline analysis}.
If a function body is not available or is too complex, the engine will evaluate the call \emph{conservatively}, invalidating the values of the variables that could have been modified by the call.
After one analysis is completed, the engine will pick a new top level function that has not been visited or inlined during the analysis of earlier top level functions.

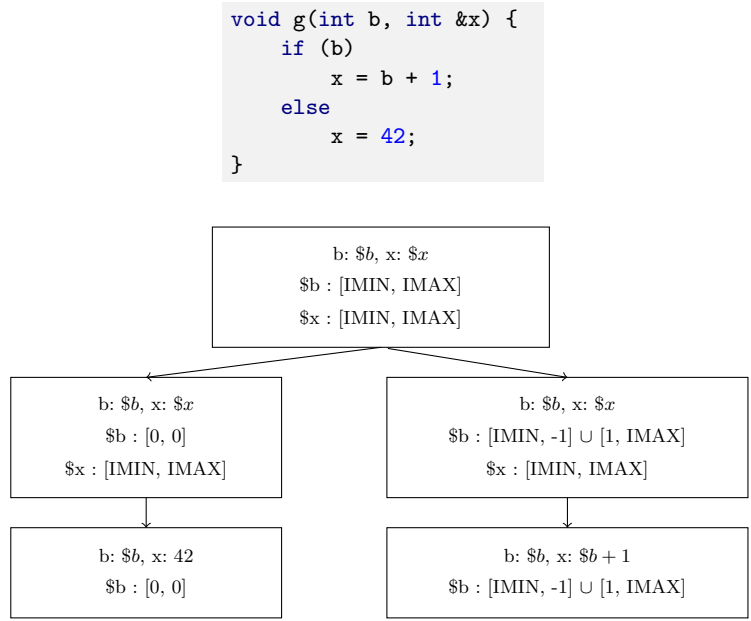
\begin{figure*}
    \centering
    \noindent\begin{minipage}{.35\textwidth}
        \begin{minted}{CPP}
            void g(int b, int &x) {
                if (b)
                    x = b + 1;
                else
                    x = 42;
            }
        \end{minted}
    \end{minipage}\vspace{10pt}
        \begin{tikzpicture}[scale=0.8, transform shape]
            \draw  (-3.4,3) rectangle (2.2,1) node[pos=.5, rectangle split,rectangle split parts=3] {b: $\$b$, x: $\$x$\nodepart{second} \$b : [IMIN, IMAX]\nodepart{third} \$x : [IMIN, IMAX]};
            \draw  (-6.75,0.5) rectangle (-2.25,-1.5) node[pos=.5, rectangle split,rectangle split parts=3] {b: $\$b$, x: $\$x$\nodepart{second} \$b : [0, 0]\nodepart{third} \$x : [IMIN, IMAX]};
            \draw  (-6.75,-2) rectangle (-2.25,-3.5) node[pos=.5, rectangle split,rectangle split parts=2] {b: $\$b$, x: $42$\nodepart{second} \$b : [0, 0]};
            \draw  (-0.5,0.5) rectangle (5.5,-1.5) node[pos=.5, rectangle split,rectangle split parts=3] {b: $\$b$, x: $\$x$\nodepart{second} \$b : [IMIN, -1]$\ \cup\ $[1, IMAX]\nodepart{third} \$x : [IMIN, IMAX]};
            \draw  (-0.5,-2) rectangle (5.5,-3.5) node[pos=.5, rectangle split,rectangle split parts=2] {b: $\$b$, x: $\$b+1$\nodepart{second} \$b : [IMIN, -1]$\ \cup\ $[1, IMAX]};
            \draw [->](-0.6,1) node (v1) {} -- (-4.5,0.5);
            \draw [->](v1) -- (2.5,0.5);
            \draw [->](-4.5,-1.5) -- (-4.5,-2);
            \draw [->](2.5,-1.5) -- (2.5,-2);
        \end{tikzpicture}
    \vspace{10pt}
    \caption{A simplified segment of the exploded graph of function \mintinline{CPP}{g()}.}\label{sa:exploded-graph}
\end{figure*}

An example analysis can be seen along with its exploded graph in \cref{sa:exploded-graph}.
The function \mintinline{CPP}{g} contains two execution paths.
Since the values of \mintinline{CPP}{b} and \mintinline{CPP}{x} are initially unknown, these values are represented with the corresponding symbols \texttt{\$b} and \texttt{\$x}.
In the beginning, these symbols can have arbitrary values.
As the analysis continues on the left execution path, the value of \texttt{b} is known to be zero.
On this same path, we later discover that the value of x is the constant \mintinline{CPP}{42}.
On the second path, the value of \mintinline{CPP}{b} can be anything but zero.
We later discover that the value of \mintinline{CPP}{x} is greater than the original value of \mintinline{CPP}{b} by one.
The symbol \texttt{\$x} is no longer needed on any of the paths, it can be garbage collected.

In the following sections we will use this framework to write our own checker module for the dangling pointer problem.

\subsection{Finding dangling pointers with symbolic execution}\label{main-section}

We saw that the exploded graph captures how the execution of a program unfolds in time.
To find a certain type of bug, we therefore need to define and recognize a series of events that lead to the erroneous situation.
In our case, that is:

\begin{enumerate}
    \setlength\itemsep{0pt}
    \item A pointer is obtained to the inner buffer of the string.
    \item An operation on the string, specified in the \CC{} standard, invalidates the pointer.
    \item The invalid pointer is used.
\end{enumerate}

A module that reports the dangling string pointer problem detailed in this chapter has been implemented in the Clang Analyzer fairly recently \linebreak
    (\texttt{cplusplus.InnerPointerChecker}~\cite{kovacs2019detecting}).
We believe that it serves as a good example to demonstrate how to write a bug-finding module of our own.

Below we summarize the most important elements of the implementation process.
Each element will be described in detail in later sections.

\paragraph{Recognizing events}

The Clang Analyzer provides a list of callbacks that run whenever a certain event is simulated in the program.
For example, implementing a \mintinline{CPP}{checkPreStmt<ASTNode>} callback will result in the engine running the contents of this function whenever it is about to evaluate a statement of \mintinline{CPP}{ASTNode} type.
In the callback, we can further specify the exact conditions we would like to match, and then perform the desired changes to the program state.
Other callbacks include matching a point where a memory location is accessed or written, the end of a function, etc.
The list of possible callbacks to implement are collected in a dedicated sample checker called \mintinline{CPP}{CheckerDocumentation}~\cite{CheckerDoc}.

\paragraph{Storing checker-specific data}

Aside from matching certain kinds of events in the code, we also need to store a list of buffer pointers for each string object we
encounter.
The bug we plan to detect occurs exactly because the language does not make this association, and the pointers can still be used after the string is destroyed.
We will store this information in a map that pairs the memory region (\mintinline{CPP}{MemRegion}) of a string object with a set of pointer symbols (\mintinline{CPP}{SymbolRef}s).
Each immutable node in the exploded graph will hold an up-to-date copy of this map.

\paragraph{Interoperability between checkers}

Exploded graph nodes also hold the data structures of other enabled checkers.
For our checker, we will always enable \texttt{unix.} \texttt{MallocChecker} as a dependency, because it already performs some of the work we need -- monitoring the usage of dangling pointers.
\mintinline{CPP}{MallocChecker} has been part of the default distribution of the Clang Analyzer for a long time, reporting the \mintinline{CPP}{malloc}/\mintinline{CPP}{free} kind of problems we described in \cref{sa:example}.
One of its data structures is the \mintinline{CPP}{RegionState} map, which stores released pointer symbols, among others.
\mintinline{CPP}{MallocChecker} then implements a list of callbacks to recognize various situations when a released pointer is used.
To reuse this functionality, our only task is to hand over pointer symbols to \mintinline{CPP}{MallocChecker} when they become invalid.

\subsection{The skeleton of the checker}

The first step towards our bug-finding module is to obtain a copy of the official LLVM repository as described in~\cite{ClangDoc}.
The initial Clang build will take somewhere between half an hour and an hour depending on our machine -- we recommend to use at least a 4-core CPU and 16 GiB of RAM if possible.
Further changes will be built incrementally, granting a reasonably fast work cycle.

Please note that any file path we reference from now on will reside inside the \texttt{llvm-project/clang/} directory.
File names will be indicated by a comment in the first line of the listing.

\subsubsection{Adding the checker to the registry}

One of our first tasks is to introduce our new module to the checker registry.
If we aim to contribute our work to the official repository, we need to be aware that new checkers are always introduced in incremental, well-reviewed steps under one of the \texttt{alpha} packages.
An \emph{alpha checker} is considered unfinished, and needs to be explicitly enabled by the user.
Checkers are pulled out of alpha after a thorough evaluation convinces reviewers that it is of high quality.
High quality in this case means that the module hardly ever produces bogus reports, but can find interesting bugs in open-source projects.

Our checker has been designed to find dangling pointer bugs connected to the \CC{} \mintinline{CPP}{std@$::$@string} class, therefore it resides in the \texttt{cplusplus} package.
(During development, it was a part of the \texttt{alpha.cplusplus} package.)
In \cref{sa:inner-ptr-reg}, we can see its checker registry entry, defining its package, name, and dependencies.

\begin{listing}[h]
    \begin{minted}{CPP}
        /* include/clang/StaticAnalyzer/Checkers/Checkers.td */

        let ParentPackage = Cplusplus in {

        def InnerPointerChecker : Checker<"InnerPointer">,
          HelpText<"Check for inner pointers of C++ containers used after "
                   "re/deallocation">,
          Dependencies<[DynamicMemoryModeling]>,
          Documentation<NotDocumented>;
    \end{minted}
    \caption{A segment of the \texttt{Checkers.td} file that lists all checker modules, classified into packages.
        Here we define \mintinline{CPP}{InnerPointerChecker} to be a part of the \texttt{cplusplus} package.}\label{sa:inner-ptr-reg}
\end{listing}

\noindent If we compile Clang with this modification, the description of our module will already be displayed by the help command (\cref{sa:help}).

\begin{listing}[h]
    \begin{minted}{text}
        $ clang -cc1 -analyzer-checker-help
        OVERVIEW: Clang Static Analyzer Checkers List

        USAGE: -analyzer-checker <CHECKER or PACKAGE,@$\ldots$@>

        CHECKERS:
          @$\ldots$@
          cplusplus.InnerPointer    Check for inner pointers of C++ containers
                                    used after re/deallocation
          @$\ldots$@
    \end{minted}
    \caption{Once a checker is added to the registry, it will be displayed by the \texttt{-analyzer-checker-help} option.}\label{sa:help}
\end{listing}

\subsubsection{Making the checker discoverable for CMake}

The file containing our checker will live in the \texttt{lib/StaticAnalyzer/Checkers/} directory, under the same name as the checker itself.
To ensure that the build system is aware of its existence, we need to include the file name in \texttt{CMakeLists.txt} (\cref{sa:cmakelists}).
We are careful to keep the alphabetical order.

\begin{listing}[h]
    \begin{minted}{CPP}
        /* lib/StaticAnalyzer/Checkers/CMakeLists.txt */

        add_clang_library(clangStaticAnalyzerCheckers
          @$\ldots$@
          IdenticalExprChecker.cpp
          InnerPointerChecker.cpp           // Our checker file.
          InvalidatedIteratorChecker.cpp
          @$\ldots$@
    \end{minted}
    \caption{The new checker file needs to be added to the \texttt{CMakeLists.txt} file.}\label{sa:cmakelists}
\end{listing}

\subsubsection{Adding the stub of the checker}

We finally add the actual file\footnote{Please note that we have edited the template to fit this article.
However, one should always use the original templates available in the repository, as the community is very strict on formal requirements.} to the \texttt{Checkers/} directory (\cref{sa:stub}).
It always starts with the LLVM license header and a short description of the module that we have left out for the sake of brevity.
For a start, we give a very basic definition of our checker class, and two functions that help register it correctly.

\begin{listing}[h!]
    \begin{minted}{CPP}
        /* lib/StaticAnalyzer/Checkers/InnerPointerChecker.cpp */

        #include "clang/StaticAnalyzer/Core/Checker.h"
        #include "clang/StaticAnalyzer/Core/CheckerManager.h"
        #include "clang/StaticAnalyzer/Core/PathSensitive/CheckerContext.h"

        using namespace clang;
        using namespace ento;

        namespace {
        class InnerPointerChecker : public Checker<> {};
        } // end of anonymous namespace

        void ento::registerInnerPointerChecker(CheckerManager &Mgr) {
          Mgr.registerChecker<InnerPointerChecker>();
        }

        bool ento::shouldRegisterInnerPointerChecker(const CheckerManager &Mgr) {
          return Mgr.getLangOpts().CPlusPlus;
        }
    \end{minted}
    \caption{Basic boilerplate for a new \CC{} checker.}\label{sa:stub}
\end{listing}

The \mintinline{CPP}{ento@$::$@shouldRegisterInnerPointerChecker} function (where the name of the given checker should be substituted) often relies on compilation options given by the user.
In this case, it only registers our module if the analysed project is built in \CC{} mode.

At this point, we should make sure that our checker stub compiles without errors.
We also need to remember that for a checker patch to be accepted to the official repository, two further changes will be needed: a file containing regression tests, and proper documentation.
Regression tests will be discussed later.

\subsection{Recognizing events}

Now that we have a basic skeleton for our checker, we can start to think about the ``business logic''.
To accomplish the first step in our plan, we need to recognize whenever one of the \mintinline{CPP}{std@$::$@string@$::$@data} or \mintinline{CPP}{std@$::$@string@$::$@c_str} methods is called.
These are the two string methods that allow the user to obtain a pointer to the inner buffer of the string.

We can recognize these events by using one of path-sensitive checker callbacks of the analyser like \mintinline{CPP}{checkPostCall}.
The engine will execute the callback right after it evaluated a function call.
The information obtained about the call is available in a \mintinline{CPP}{CallEvent} object.
We can check whether one of the target functions was called by defining \mintinline{CPP}{CallDescription} objects for them, and passing them to the \mintinline{CPP}{CallEvent@$::$@isCalled} method (\cref{sa:callback}).

\begin{listing}[h]
    \begin{minted}{CPP}
        #include "clang/StaticAnalyzer/Core/PathSensitive/CallEvent.h"

        class InnerPointer : public Checker<check::PostCall> {
          CallDescription CStrFn, DataFn;

        public:
          InnerPointerChecker() :
            CStrFn({"std", "basic_string", "c_str"}),
            DataFn({"std", "basic_string", "data"}) {}

          void checkPostCall(const CallEvent &Call, CheckerContext &C) const;
        };

        void InnerPointerChecker::checkPostCall(const CallEvent &Call,
                                                CheckerContext &C) const {
          if (Call.isCalled(CStrFn) || Call.isCalled(DataFn)) {
            // $\ldots$
          }
        }
    \end{minted}
    \caption{Extending our checker to match \mintinline{CPP}{data} and \mintinline{CPP}{c_str} calls.}\label{sa:callback}
\end{listing}

\subsubsection{Storing checker-specific data}

Whenever a buffer pointer obtaining method is called on a string, the analyser models it by returning a new symbol representing a new pointer.
This is called a \emph{conservative evaluation} of a function call, and it often happens when the engine cannot understand the complicated templated code of a standard library within a reasonable budget.
If this is the case, it safely assumes that the call may return a different result each time it is called.

This means that we will have to store a whole list of symbols for each string object, corresponding to the results of different \mintinline{CPP}{c_str()} and \mintinline{CPP}{data()} calls on it.
To create a new map data structure in the program state, we can use the \mintinline{CPP}{REGISTER_MAP_WITH_PROGRAMSTATE} macro on the top of our file, in the global namespace.
Its name will be \mintinline{CPP}{RawPtrMap}, and it will map the \mintinline{CPP}{MemRegion}-type memory location of a string object to a set of \mintinline{CPP}{SymbolRef}s that represent the pointers (\cref{sa:register}).

\begin{listing}[h!]
    \begin{minted}{CPP}
        // Associate container objects with a set of raw pointer symbols.
        REGISTER_SET_FACTORY_WITH_PROGRAMSTATE(PtrSet, SymbolRef)
        REGISTER_MAP_WITH_PROGRAMSTATE(RawPtrMap, const MemRegion*, PtrSet)
    \end{minted}
    \caption{New data structures can be added to the program state using different \mintinline{CPP}{REGISTER_(@$\ldots$@)_WITH_PROGRAMSTATE} macros.
        We added these lines to the global namespace, just before our class.}\label{sa:register}
\end{listing}

We also need to define how our set type will be used.
If we wanted to create a stand-alone set in the program state, just like our map, then we would use the \mintinline{CPP}{REGISTER_SET_WITH_PROGRAMSTATE} macro.
However, we do not wish to use this set individually, we would only use it as part of the map structure.
We will create and modify instances of our set though its factory, which can be declared by the \mintinline{CPP}{REGISTER_SET_FACTORY_WITH_PROGRAMSTATE} macro.
Similar available program state traits can be created using the macros defined in \texttt{ProgramStateTrait.h}.

\begin{listing}[h]
    \begin{minted}[linenos]{CPP}
    if (const auto *ICall = dyn_cast<CXXInstanceCall>(&Call)) {
      const MemRegion *String = ICall->getCXXThisVal().getAsRegion();
      if (!String)
        return;

      if (Call.isCalled(CStrFn) || Call.isCalled(DataFn)) {
        SVal RawPtr = Call.getReturnValue();
        if (SymbolRef Sym = RawPtr.getAsSymbol(true)) {
          ProgramStateRef State = C.getState();
          ProgramStateManager &PSM = State->getStateManager();
          PtrSet@$::$@Factory &F = PSM.get_context<PtrSet>();

          const PtrSet *SetPtr = State->get<RawPtrMap>(String);
          PtrSet Set = SetPtr ? *SetPtr : F.getEmptySet();
          Set = F.add(Set, Sym);

          State = State->set<RawPtrMap>(String, Set); @$\label{lin:rawptrmap}$@
          C.addTransition(State);
        }
        return;
      }
    }
    \end{minted}
    \caption{Updating the set of pointer symbols corresponding to a given string object is a little cumbersome.
        At least obtaining the \mintinline{CPP}{this} value and return value of a method call is straightforward.
        In the end, we create a new exploded graph node holding the modified program state.}\label{sa:post-call}
\end{listing}

We can now continue implementing our \mintinline{CPP}{checkPostCall} callback (\cref{sa:post-call}).
We first obtain the \mintinline{CPP}{MemRegion} of the string object for our map entry.
It can easily be retrieved as the \mintinline{CPP}{this} value of the examined call - however, that only exists if the call is a \CC{} method call (here: \mintinline{CPP}{CXXInstanceCall}).
(For the methods matching nodes in AST, see \cref{tidy:matching}.)
The \mintinline{CPP}{CallEvent} type hierarchy can be viewed on its documentation page~\cite{CallEvent}.
We do not need to check whether this \mintinline{CPP}{MemRegion} is a string's \mintinline{CPP}{MemRegion}, because the \mintinline{CPP}{CallDescription} checks that follow make it clear.

We first obtain the return value of the method call from the \mintinline{CPP}{CallEvent}.
\mintinline{CPP}{SVal}~\cite{SVal} is the most generic symbolic value type inside the analyser.
It can hold values that can be converted to \mintinline{CPP}{MemRegion} type (representing a memory region of an object), to \mintinline{CPP}{SymbolRef} type (representing symbolic values that appear in, e.g.\ branching conditions), or values that cannot be converted to either of them.

We expect that the result of the examined method call can be converted to a \mintinline{CPP}{SymbolRef}, because pointers and integers are usually of this type.
We obtain the factory for our set type created earlier (\mintinline{CPP}{PtrSet}), and see if an instance has already been created for the given string object.
If not, we now create a new empty set.
We then add the returned pointer symbol to the set.

At this point, we have created a new, updated set of buffer pointers for our string, but we have not committed this information back to \mintinline{CPP}{RawPtrMap} yet.
We do it on Line~\ref{lin:rawptrmap} in \cref{sa:post-call}.

The \mintinline{CPP}{addTransition} call on \mintinline{CPP}{CheckerContext} is important.
Remember that the program state is an immutable data structure.
Whenever something changes in the program state, a new exploded graph node needs to be created by an \mintinline{CPP}{addTransition} call.
The new node we create here is a copy of the previous node, apart from the modifications we made to \mintinline{CPP}{RawPtrMap}.
It is advised to write a return statement after each \mintinline{CPP}{addTransition} call, to ensure that no unintended new nodes are created.

\subsection{Pointer invalidation}

So far, we have started tracking the buffer symbols of strings.
Next, we will need to recognize when they become invalidated.
For this purpose, we reference the \texttt{[string.require]} section of the \CC{} standard~\cite{StdDraft}:

\begin{displayquote}
    \small{``References, pointers, and iterators referring to the elements of a \mintinline{CPP}{basic_string} sequence may be invalidated by the following uses of that \mintinline{CPP}{basic_string} object:}

    \vspace{5pt}

    -- As an argument to any standard library function taking a reference to non-const \mintinline{CPP}{basic_string} as an argument.

    \vspace{5pt}

    -- Calling non-const member functions, except \mintinline{CPP}{operator[]}, \mintinline{CPP}{at}, \mintinline{CPP}{front}, \mintinline{CPP}{back}, \mintinline{CPP}{begin}, \mintinline{CPP}{rbegin}, \mintinline{CPP}{end}, and \mintinline{CPP}{rend}.''
\end{displayquote}

\noindent Because buffer pointers become invalid after the above described function calls take place, we can continue implementing this feature in the \mintinline{CPP}{checkPostCall} function we have started.

\subsubsection{Invalidation by member call}

Let us first deal with the second point, concerning non-const member functions, as we already have a conditional in place for member calls.
Supposing that we have a function that checks whether the given call invalidates buffer pointers (\mintinline{CPP}{isInvalidatingMemberFunction}), our task is to mark all the pointer symbols associated with the string \emph{released} (done by \mintinline{CPP}{markPtrSymbolsReleased}).
See \cref{sa:inv-mem-func}.

\begin{listing}[h]
    \begin{minted}{CPP}
        ProgramStateRef State = C.getState();

        if (const auto *ICall = dyn_cast<CXXInstanceCall>(&Call)) {
          const MemRegion *String = ICall->getCXXThisVal().getAsRegion();
          if (!String)
            return;

          if (Call.isCalled(CStrFn) || Call.isCalled(DataFn)) {
            // $\ldots$
          }

          if (isInvalidatingMemberFunction(Call)) {
            markPtrSymbolsReleased(Call, State, String, C);
            return;
          }
        }
    \end{minted}
    \caption{Pointer invalidation logic can be added right after our previous work in \mintinline{CPP}{checkPostCall}.}\label{sa:inv-mem-func}
\end{listing}

\noindent Let us return to this function later, and now focus on the implementation of \mintinline{CPP}{isInvalidatingMemberFunction}.

First, we define a \mintinline{CPP}{CallDescription} object for each non-const member function of \mintinline{CPP}{std@$::$@string} that is not excluded in the second point (\cref{sa:inv-mem-func-2}).

\begin{listing}[h!]
    \begin{minted}{CPP}
        class InnerPointerChecker : public Checker<check::PostCall> {

          CallDescription AppendFn, AssignFn, ClearFn, CStrFn, DataFn, EraseFn,
              InsertFn, PopBackFn, PushBackFn, ReplaceFn, ReserveFn, ResizeFn,
              ShrinkToFitFn, SwapFn;

        public:
          InnerPointerChecker()
              : AppendFn({"std", "basic_string", "append"}),
                // $\ldots$
                SwapFn({"std", "basic_string", "swap"}) {}
    \end{minted}
    \caption{Defining \mintinline{CPP}{CallDescription}s for the list of non-const member functions of \mintinline{CPP}{std@$::$@string} specified in the standard.
        We have omitted repetitive definitions for the sake of brevity.}\label{sa:inv-mem-func-2}
\end{listing}

Some special member functions (destructor, member operators) cannot be easily represented by a \mintinline{CPP}{CallDescription} at the moment, they need to be checked by hand.
Our implementation of \mintinline{CPP}{isInvalidatingMemberFunction} can bee seen in \cref{sa:inv-mem-func-3}.

\begin{listing}[h!]
    \begin{minted}{CPP}
        static bool isInvalidatingMemberFunction(const CallEvent &Call) {
          if (const auto *MOC = dyn_cast<CXXMemberOperatorCall>(&Call)) {
            OverloadedOperatorKind Opc = MOC->getOverloadedOperator();
            return Opc == OO_Equal || Opc == OO_PlusEqual;
          }
          return (isa<CXXDestructorCall>(Call) || Call.isCalled(AppendFn)
                 || Call.isCalled(AssignFn)    || Call.isCalled(ClearFn)
                 || Call.isCalled(EraseFn)     || Call.isCalled(InsertFn)
                 || Call.isCalled(PopBackFn)   || Call.isCalled(PushBackFn)
                 || Call.isCalled(ReplaceFn)   || Call.isCalled(ReserveFn)
                 || Call.isCalled(ResizeFn)    || Call.isCalled(ShrinkToFitFn)
                 || Call.isCalled(SwapFn));
        }
    \end{minted}
    \caption{Pointers to a string are invalidated by calling non-const member functions on the string, except \mintinline{CPP}{operator[]}, \mintinline{CPP}{at}, \mintinline{CPP}{front}, \mintinline{CPP}{back}, \mintinline{CPP}{begin}, \mintinline{CPP}{rbegin}, \mintinline{CPP}{end}, and \mintinline{CPP}{rend}.}\label{sa:inv-mem-func-3}
\end{listing}

\subsubsection{Invalidation by library function call}

The other point in the standard states that pointers to a string may also become invalid if the string is passed to a standard library function by non-const reference.
In this case, we are interested in all kinds of functions, so we place our logic outside of the member call conditional, at the end of \mintinline{CPP}{checkPostCall} (\cref{sa:inv-func-use}).

\begin{listing}[h]
    \begin{minted}{CPP}
        if (const auto *ICall = dyn_cast<CXXInstanceCall>(&Call)) {
          // $\ldots$
        }

        checkFunctionArguments(Call, State, C);
    \end{minted}
    \caption{We place our logic for checking library function arguments outside of the instance call conditional, at the end of \mintinline{CPP}{checkPostCall}.}\label{sa:inv-func-use}
\end{listing}

\begin{listing}[h!]
    \begin{minted}{CPP}
        static void checkFunctionArguments(const CallEvent &Call,
                                           ProgramStateRef State,
                                           CheckerContext &C) {
          if (const auto *FC = dyn_cast<AnyFunctionCall>(&Call)) {
            const FunctionDecl *FD = FC->getDecl();
            if (!FD || !FD->isInStdNamespace())
              return;

            for (unsigned I = 0, E = FD->getNumParams(); I != E; ++I) {
              QualType ParamTy = FD->getParamDecl(I)->getType();
              if (!ParamTy->isReferenceType() ||
                  ParamTy->getPointeeType().isConstQualified())
                continue;

              // In case of member operator calls, `this` is counted
              // as an argument but not as a parameter.
              bool isaMemberOpCall = isa<CXXMemberOperatorCall>(FC);
              unsigned ArgI = isaMemberOpCall ? I+1 : I;

              SVal Arg = FC->getArgSVal(ArgI);
              const MemRegion *ArgRegion = Arg.getAsRegion();
              if (!ArgRegion)
                continue;

              markPtrSymbolsReleased(Call, State, ArgRegion, C);
            }
          }
        }
    \end{minted}
    \caption{Buffer pointers may be invalidated if the string is passed as an argument to any standard library function that takes a reference to non-const string as an argument.}\label{sa:inv-func-def}
\end{listing}

We define this function as a static utility function outside of the checker class, similarly to \mintinline{CPP}{isInvalidatingMemberFunction} (\cref{sa:inv-func-def}).
In the beginning, we obtain the declaration of the called function.
We then iterate over the parameters of the function, processing only those that are taken by non-const reference.
Next, we are interested in the actual arguments the function received.
In most cases, converting a parameter index to the corresponding argument index is straightforward, but we need to be careful with member operators, where \emph{this} is not counted as a parameter.
We then obtain the symbolic value associated with the argument at this point of the simulation.
(Remember that \mintinline{CPP}{checkPostCall} runs after a call is evaluated by the engine.)
If it represents a memory location, it is a good candidate to be one of the strings we track.
We delegate the tasks of looking up this region in the program state map and performing the invalidation to \mintinline{CPP}{markPtrSymbolsReleased}.

\subsubsection{Performing the invalidation}

At this point, we are able to recognize any operation on a string that invalidates its buffer pointers.
Our next task is to modify our data structures to reflect this change.

As described in the beginning of \cref{main-section}, the usage of released pointers is already being monitored by a module called \mintinline{CPP}{MallocChecker}.
\mintinline{CPP}{MallocChecker} maintains a list of released pointer symbols, and subscribes to various events in the program that may involve the use of pointers.
If any event involves the use of a released pointer, the checker reports the error to the user, as seen in \cref{sa:use-after-free-c}.

For our \mintinline{CPP}{markPtrSymbolsReleased} function, we have a memory region that potentially represents a string we track.
We start by checking whether it is present in our program state map.
If it is not, then nobody has ever asked for a buffer pointer for that string (or indeed it is not a string at all), and we have no work to do.
If the region is present in the map, however, we want to iterate over all the pointer symbols associated with it, and hand them over to \mintinline{CPP}{MallocChecker} for usage tracking.
After that, we consider the string to be different, and can thus remove the whole entry from the program state map.
The implementation can be seen in \cref{sa:mark-invalid}.

\begin{listing}[h!]
    \begin{minted}{CPP}
        static void markPtrSymbolsReleased(const CallEvent &Call,
                                           ProgramStateRef State,
                                           const MemRegion *MR,
                                           CheckerContext &C) {
          if (const PtrSet *PS = State->get<RawPtrMap>(MR)) {
            const Expr *Orig = Call.getOriginExpr();
            for (const auto Symbol : *PS) {
              // `Orig` may be null, and will be stored so in the symbol's
              // `RefState` in MallocChecker's `RegionState` map.
              State = allocation_state::markReleased(State, Symbol, Orig);
            }
            State = State->remove<RawPtrMap>(MR);
            C.addTransition(State);
            return;
          }
        }
    \end{minted}
    \caption{This function checks whether the given memory region represents a string we track.
    If so, it sends the associated buffer pointer symbols over to \mintinline{CPP}{MallocChecker} in a \emph{released} state.}\label{sa:mark-invalid}
\end{listing}

We have used a function named \mintinline{CPP}{markReleased} to pass over one given symbol to \mintinline{CPP}{MallocChecker}, together with the call expression from which it originates.
The latter is not important in our case, but \mintinline{CPP}{MallocChecker} always stores the origin expression in its \mintinline{CPP}{RegionState} map.

Let us now define this function, showcasing one possible way of cooperation between checkers.
We suspect that we will need a few more shared functions for bug reporting, so it makes sense to create a new header that we call \texttt{AllocationState.h} (\cref{sa:alloc-state-header}).
We can create this file right next to the checker definition files in the \texttt{lib/StaticAnalyzer/Checkers/} folder.

\begin{listing}[h!]
    \begin{minted}{CPP}
        #ifndef LLVM_CLANG_LIB_STATICANALYZER_CHECKERS_ALLOCATIONSTATE_H
        #define LLVM_CLANG_LIB_STATICANALYZER_CHECKERS_ALLOCATIONSTATE_H

        #include "clang/StaticAnalyzer/Core/PathSensitive/ProgramState.h"

        namespace clang {
        namespace ento {
        namespace allocation_state {

        ProgramStateRef markReleased(ProgramStateRef State, SymbolRef Sym,
                                     const Expr *Origin);

        } // end namespace allocation_state
        } // end namespace ento
        } // end namespace clang

        #endif
    \end{minted}
    \caption{This header file declares a function that provides interoperability between \mintinline{CPP}{InnerPointerChecker} and \mintinline{CPP}{MallocChecker}.
        It is called in the former, but implemented in the latter.}\label{sa:alloc-state-header}
\end{listing}

It is important that we return a reference to a program state from this function.
\mintinline{CPP}{MallocChecker} will create a new copy of the program state, adding the received pointer symbol to its own data structure that holds a list of released symbols.
However, this new state only becomes a part of the exploded graph once \mintinline{CPP}{addTransition} is called in \mintinline{CPP}{markPtrSymbolsReleased}.
See the implementation in \cref{sa:alloc-state-imp}.

\begin{listing}[h!]
    \begin{minted}{CPP}
        namespace clang {
        namespace ento {
        namespace allocation_state {

        ProgramStateRef markReleased(ProgramStateRef State, SymbolRef Sym,
                                     const Expr *Origin) {
          AllocationFamily Family = AF_InnerBuffer;
          return State->set<RegionState>(Sym,
                                         RefState::getReleased(Family, Origin));
        }

        } // end namespace allocation_state
        } // end namespace ento
        } // end namespace clang
    \end{minted}
    \caption{The implementation of the function declared in \texttt{AllocationState.h} as implemented in \texttt{MallocChecker.cpp}.}\label{sa:alloc-state-imp}
\end{listing}

\mintinline{CPP}{AllocationFamily} is an internal class in \mintinline{CPP}{MallocChecker} to distinguish between different sources of symbols.
We have created a new family to represent pointers stemming from \CC{} containers called \mintinline{CPP}{AF_InnerBuffer}.
We omit the details of this addition for the sake of brevity.
\mintinline{CPP}{RefState} is \mintinline{CPP}{MallocChecker}'s private type to describe the allocation state of a pointer symbol.
Its \mintinline{CPP}{getReleased} method constructs an appropriate instance of this type to be saved together with the pointer in the \mintinline{CPP}{RegionState} map.

With these additions in place, we can rely on \mintinline{CPP}{MallocChecker} to find the uses of released string buffer pointers for us.

\subsection{The bug report}

\subsubsection{The warning message}

We concluded the previous section by handing over dangling string buffer pointers to \mintinline{CPP}{MallocChecker}.
\mintinline{CPP}{MallocChecker} will subscribe to various situations in the program when a pointer variable is used.
It will check whether the used pointer symbol is among the released symbols it tracks - if so, it will call a function named \mintinline{CPP}{HandleUseAfterFree} (see \cref{sa:handle-use-after-free}).

\begin{listing}[h!]
    \begin{minted}{CPP}
        void MallocChecker::HandleUseAfterFree(CheckerContext &C,
                                               SourceRange Range,
                                               SymbolRef Sym) const {
          // $\ldots$

          if (ExplodedNode *N = C.generateErrorNode()) {
            if (!BT_UseFree[*CheckKind])
              BT_UseFree[*CheckKind].reset(new BugType(
                  CheckNames[*CheckKind], "Use-after-free",
                  categories::MemoryError));

            AllocationFamily AF =
                C.getState()->get<RegionState>(Sym)->getAllocationFamily();

            auto R = std::make_unique<PathSensitiveBugReport>(
                *BT_UseFree[*CheckKind],
                AF == AF_InnerBuffer
                    ? "Inner pointer of container used after re/deallocation"
                    : "Use of memory after it is freed",
                N);

            R->addRange(Range);
            R->addVisitor(std::make_unique<MallocBugVisitor>(Sym));

            if (AF == AF_InnerBuffer)
              R->addVisitor(allocation_state::getInnerPointerBRVisitor(Sym));

            C.emitReport(std::move(R));
          }
        }
    \end{minted}
    \caption{This function creates the use-after-free warning for the dangling pointer problem.
      Less interesting parts have been edited out for brevity.}\label{sa:handle-use-after-free}
\end{listing}

The function begins by generating a new \emph{error node} to be included in the exploded graph.
This special kind of node will be attached to the path-sensitive bug report.
As we have detected a critical memory error, the analysis will not proceed any further on this path.

\mintinline{CPP}{BT_UseFree} is a private array of \mintinline{CPP}{std@$::$@unique_ptr<BugType>} type objects, one for each kind of check that \mintinline{CPP}{MallocChecker} supports.
We initialize the slot belonging to our own checker if needed.
These bug types are used to categorize reports when users view them.

Then, we create a new \mintinline{CPP}{PathSensitiveBugReport} object, attaching the appropriate bug type, a custom error message, and the error node itself.
We include the \mintinline{CPP}{SourceRange} object that represents the location of the error in the source text, and two different \emph{bug reporter visitors}.
Finally, we hand over the carefully assembled report object to \mintinline{CPP}{CheckerContext}.

\subsubsection{A note for the point of release}

When the analysis is finished and bug reports are collected, bug reporter visitors perform further processing on the reports, attaching explaining notes to the bug paths.

\mintinline{CPP}{MallocChecker} has its own visitor called \mintinline{CPP}{MallocBugVisitor}, defined in the same file as the checker-complex itself.
The most interesting part of the visitor is its \mintinline{CPP}{VisitNode} function.
Starting from the error node and going backwards, this function is called on each node of the bug path.

We have enhanced the \mintinline{CPP}{MallocBugVisitor::VisitNode} function to attach a custom explaining note to the moment when the string buffer pointer became released.
This event is identified by a change in the \mintinline{CPP}{RegionState} map -- proceeding backwards on the path, starting from the error node, we are looking for the last node that contains the pointer as a released symbol.
Then we assemble the message that includes the type of the container (string, in our case), and the name of the called method that caused the release.
A small and edited segment of the changes can be seen in \cref{sa:malloc-visit-node}.

\begin{listing}[h!]
    \begin{minted}[breaklines]{CPP}
          auto *ObjR = allocation_state::getContainerObjRegion(statePrev, Sym);
          QualType ObjTy = cast<TypedValueRegion>(ObjR)->getValueType();
          OS << "Inner buffer of '" << ObjTy.getAsString() << "' ";

          if (N->getLocation().getKind() == ProgramPoint::PostImplicitCallKind) {
            OS << "deallocated by call to destructor";
          } else {
            OS << "reallocated by call to '";
            const Stmt *S = RSCurr->getStmt();
            if (const auto *MemCallE = dyn_cast<CXXMemberCallExpr>(S)) {
              OS << MemCallE->getMethodDecl()->getDeclName();
            }
            else if (const auto *OpCallE = dyn_cast<CXXOperatorCallExpr>(S)) {
              OS << OpCallE->getDirectCallee()->getDeclName();
            }
            else if (const auto *CallE = dyn_cast<CallExpr>(S)) {
              auto &CEMgr = BRC.getStateManager().getCallEventManager();
              CallEventRef<> Call = CEMgr.getSimpleCall(CallE, state, CurrentLC);
              if (const auto *D = dyn_cast_or_null<NamedDecl>(Call->getDecl()))
                OS << D->getDeclName();
              else
                OS << "unknown";
            }
            OS << "'";
          }
          Msg = OS.str();
    \end{minted}
    \caption{A heavily edited segment of \mintinline{CPP}{MallocBugVisitor::VisitNode}.
        This part assembles an explaining note to be attached to the point in the bug path when the string buffer was freed.
    }\label{sa:malloc-visit-node}
\end{listing}

The \mintinline{CPP}{MemRegion} of the container is obtained via a function that we define in \mintinline{CPP}{InnerPointerChecker}.
Because it was the responsibility of this checker to monitor the strings and its buffer pointers, it is the only entity that has access to this information.
We do not show this function, as it is a simple lookup in the \mintinline{CPP}{RawPtrMap} recorded in the given program state for the given pointer symbol.
For easy interoperability, we have declared it in the same header and namespace as we did with the \mintinline{CPP}{markReleased} function.
The type of the container can then be obtained in a string format easily.

For the freeing method call, the automatic destructor is a special case.
It is not a conventional function call, but is best indicated by a special kind of program point called \mintinline{CPP}{PostImplicitCallKind}.
We differentiate the other method calls by their expression types, attempting to obtain the best possible name from their declaration.

\subsubsection{A note for the pointer acquisition}

There is another point along the bug path that users might be interested in.
The moment when they obtained the buffer pointer from the string will tell them which pointer caused the error.
This moment was handled by \mintinline{CPP}{InnerPointerChecker}, therefore it is its responsibility to add the note.

This checker class does not have a bug reporter visitor that we can extend, so we need to create our own  (\cref{sa:inner-BR}).
We may add this as a nested class inside our checker class.
It needs to be a subclass of \mintinline{CPP}{BugReporterVisitor}, and it will have a \mintinline{CPP}{SymbolRef} field to store our dangling pointer symbol.

\begin{listing}[h!]
    \begin{minted}[breaklines]{CPP}
  class InnerPointerBRVisitor : public BugReporterVisitor {
    SymbolRef PtrToBuf;

  public:
    InnerPointerBRVisitor(SymbolRef Sym) : PtrToBuf(Sym) {}

    virtual PathDiagnosticPieceRef
    VisitNode(const ExplodedNode *N, BugReporterContext &BRC,
              PathSensitiveBugReport &BR) override;
  };
    \end{minted}
    \caption{A basic definition of our bug reporter visitor class.
    }\label{sa:inner-BR}
\end{listing}

The implementation of the \mintinline{CPP}{VisitNode} function can be seen in \cref{sa:inner-BR-visitnode}.
It works the same way as in \mintinline{CPP}{MallocChecker}: starting from the error node, walking backwards on the bug path, this function gets called with each exploded node.
We are looking for the moment when we started tracking the pointer symbol that later became dangling.
This moment is characterized by a change in \mintinline{CPP}{RawPtrMap}: at one program point the symbol was not yet present in the map, but in the next one it is.
For this reason we are always considering pairs of adjacent exploded nodes: one is the current node \mintinline{CPP}{N} given as an argument to the function, the other is its direct predecessor  in the exploded graph, which can be obtained by \mintinline{CPP}{N->getFirstPred()}.

The main logic is exercised by the first conditional.
\mintinline{CPP}{isSymbolTracked} is a utility method we have added to the \mintinline{CPP}{InnerPointerBRVisitor} class that we omit from this paper for the sake of brevity.
It does nothing else than a lookup in \mintinline{CPP}{RawPtrMap} as recorded at the program state provided by the first argument, for the symbol provided by the second argument.
We wish to express that the given symbol was tracked in the current exploded node, but not in its first predecessor node:

\vspace{5pt}
\mintinline{CPP}{isSymbolTracked(ThisState) && !isSymbolTracked(PredState)}
\vspace{5pt}

\noindent Because the Coding Standards of the LLVM project encourage early exits instead of nested conditionals, we will exit the function early if the negation of this  expression is true, as can be seen in \cref{sa:inner-BR-visitnode}.

If we have found the appropriate node, however, we proceed to assemble a new \mintinline{CPP}{PathDiagnosticEventPiece}, which holds all the information needed  to explain the situation to the user: the position in the source code and a short message describing the container from which the pointer originates.
We obtain the type of the container by using the same  auxiliary function we have used in \mintinline{CPP}{MallocChecker}'s visitor (\mintinline{CPP}{getContainerObjRegion}).

\begin{listing}[h!]
    \begin{minted}[breaklines]{CPP}
PathDiagnosticPieceRef
InnerPointerChecker::InnerPointerBRVisitor::VisitNode(
    const ExplodedNode *N, BugReporterContext &BRC,
    PathSensitiveBugReport &) {

  if (!isSymbolTracked(N->getState(), PtrToBuf) ||
      isSymbolTracked(N->getFirstPred()->getState(), PtrToBuf))
    return nullptr;

  const Stmt *S = N->getStmtForDiagnostics();
  if (!S)
    return nullptr;

  const MemRegion *ObjRegion =
      allocation_state::getContainerObjRegion(N->getState(), PtrToBuf);
  const auto *TypedRegion = cast<TypedValueRegion>(ObjRegion);
  QualType ObjTy = TypedRegion->getValueType();

  SmallString<256> Buf;
  llvm::raw_svector_ostream OS(Buf);
  OS << "Pointer to inner buffer of '" << ObjTy.getAsString()
     << "' obtained here";
  PathDiagnosticLocation Pos(S, BRC.getSourceManager(),
                             N->getLocationContext());
  return std::make_shared<PathDiagnosticEventPiece>(
    Pos, OS.str(), true);
}
    \end{minted}
    \caption{\mintinline{CPP}{InnerPointerChecker}'s visitor adds an explaining note to the point where the raw pointer symbol (that later becomes dangling) was obtained from the string object.}\label{sa:inner-BR-visitnode}
\end{listing}

\subsection{Regression tests}

Although we have left this topic to the end of the chapter, the author of a new module should always accompany their changes with appropriate tests, at each step of the process.
If the changes do not concern the analyser engine, then this means adding regression tests for the checker to a new file created in the \texttt{test/Analysis} directory.
A small segment of the test file for \mintinline{CPP}{InnerPointerChecker} can be seen in \cref{sa:test}.

\begin{listing}[h!]
    \begin{minted}[breaklines]{CPP}
        /* clang/test/Analysis/inner-pointer.cpp */

        // RUN: %clang_analyze_cc1 -std=c++14
        // RUN:   -analyzer-checker=cplusplus.InnerPointer \
        // RUN:   %s -analyzer-output=text -verify

        void deref_after_clear() {
          const char *c;
          std@$::$@string s;
          c = s.c_str(); // expected-note {{Pointer to inner buffer of 'std$::$string' obtained here}}
          s.clear();     // expected-note {{Inner buffer of 'std$::$string' reallocated by call to 'clear'}}
          consume(c);
          // expected-warning@-1 {{Inner pointer of container used after re/deallocation}}
          // expected-note@-2 {{Inner pointer of container used after re/deallocation}}
        }
    \end{minted}
    \caption{A small section of the regression test file for \mintinline{CPP}{InnerPointerChecker}.}\label{sa:test}
\end{listing}

Regression testing is managed by the LLVM Integrated Tester tool called \texttt{lit}.
Parameters are specified in the \mintinline{CPP}{RUN} lines commented out at the beginning of the file.
\texttt{\%clang\_analyze\_cc1} is a common abbreviation to run the compiler frontend with the analysis engine enabled.
Individual checkers or checker groups can be specified with the \mintinline{CPP}{-analyzer-checker} option.
The two most important settings for the testing of our checker are \mintinline{CPP}{-analyzer-output=text} and \mintinline{CPP}{-verify}.
The first setting enables notes to be displayed on the command line (by default only warnings are printed), and \mintinline{CPP}{-verify} checks that textual output against the directives specified in the test file itself.

Verifying directives can be added in the form of comments on appropriate source code lines.
\texttt{expected-warning} specifies the warning message that should be displayed on the given line, \texttt{expected-note} is the same for notes.
(Due to quirks in the engine, an identical note is printed for the warning as well.)
If the directive is placed on a different line that its target source code line, then the \texttt{@-N} and \texttt{@+N} modifiers can be used to give the distance (\texttt{N}) in lines.

\subsection{Summary}

We have walked through the implementation of a path-sensitive static analysis module in the Clang Static Analyzer framework in this section.
We first explored the dangling pointer problem as a motivation for our work, and discussed why path-sensitive analysis is more fitting for this purpose than plain AST-based analysis.
We learned the basics of the path-sensitive analysis technique, \emph{symbolic execution}, that the Analyzer uses.
Then we outlined the fundamental steps our analysis needed to take, and shown the source code of an implementation.

The problem we decided to solve in this chapter is special in a sense that we could re-use some of the functionality already present in the analysis toolset.
In turn, our task was more complicated in that we had to build out the infrastructure to accommodate the communication between the existing modules and our own.
Most checker ideas can probably be solved by using a single checker class, so we encourage the Reader to try it out for themselves.

\clearpage
\section{Conclusion}\label{conclusion}

Static analysis methods evaluate software systems based on their source code without running them.
As static analysis does not require to choose specific input values or even a working run-time environment it is much easier to integrate it to the Continuous Integration loop.
Although static analysis has its theoretical limitations -- and the applied heuristics may overestimate or underestimate the programs -- in practice it proved to be effective in analysing industrial scale systems.

Various analysis methods exists and are supported by different analysis frameworks.
The Abstract syntax tree-based methods are usually faster and require less resources during execution, but do not provide path sensitive analysis.
Symbolic execution is capable for more precise analysis as it is based on abstract interpretation but on the cost of higher resource requirements.
Both of these methods could find their role in modern software development.

In this tutorial we presented two real world \CC{} problems.
One of them is related with the elimination of redundant pointers and was solved by an AST based checker, the other is dealing the problem of dangling pointers originated from the \mintinline{CPP}{std@$::$@string} class.
We explained the different capabilities of the checkers and provided a step by step guide for their implementation.
Following these steps one can learn the technical background and the design method of the different approaches and can start to contribute to the open source word of static analysis.
For those developers, who are the users of the static analysis tools, this tutorial helps to understand the advantages and limitations of the individual methods.

\section*{Acknowledgements}
This work is a result of the implementation of the Erasmus{\small{}+} Key Action $2$ project \textnumero{} \texttt{2017-1-SK01-KA203-035402}: \emph{``Focusing Education on Composability, Comprehensibility, and Correctness of Working Software''}.

%
%
%
\bibliographystyle{splncs04}
\bibliography{bibliography}

\begin{thebibliography}{10}
\providecommand{\url}[1]{\texttt{#1}}
\providecommand{\urlprefix}{URL }
\providecommand{\doi}[1]{https://doi.org/#1}

\bibitem{aho1975deterministic}
Aho, A.V., Johnson, S.C., Ullman, J.D.: Deterministic parsing of ambiguous
  grammars. Communications of the ACM  \textbf{18}(8),  441--452 (1975)

\bibitem{AppleAnalyze}
{Apple Inc.}: Analyzing Your Code (2016),
  \url{http://developer.apple.com/library/archive/documentation/ToolsLanguages/Conceptual/Xcode_Overview/AnalyzingYourCode.html},
  (accessed 2020.12.21.)

\bibitem{bacon1996fast}
Bacon, D.F., Sweeney, P.F.: Fast static analysis of c++ virtual function calls.
  In: Proceedings of the 11th ACM SIGPLAN conference on Object-oriented
  programming, systems, languages, and applications. pp. 324--341 (1996)

\bibitem{barendregt1992lambda}
Barendregt, H.P.: Lambda calculi with types  (1992)

\bibitem{Bessey2010}
Bessey, A., Block, K., Chelf, B., Chou, A., Fulton, B., Hallem, S., Henri-Gros,
  C., Kamsky, A., McPeak, S., Engler, D.: A few billion lines of code later:
  Using static analysis to find bugs in the real world. Commun. ACM
  \textbf{53}(2),  66--75 (Feb 2010). \doi{10.1145/1646353.1646374},
  \url{http://doi.acm.org/10.1145/1646353.1646374}

\bibitem{fixcost}
Boehm, B., Basili, V.R.: Software defect reduction top 10 list. Computer
  \textbf{34}(1),  135--137 (1 2001). \doi{10.1109/2.962984},
  \url{http://dx.doi.org/10.1109/2.962984}

\bibitem{BrunnerICAI2017}
Brunner, T., Porkol\'{a}b, Z.: Programming language history: Experiences based
  on the evolution of {\CC{}}. In: Kusper, G., Kir{\'{a}}ly, R., Kunkli, R.,
  Kir\'{a}ly, S., Tibor, T. (eds.) {P}roceedings of the 10th {I}nternational
  {C}onference on {A}pplied {I}nformatics. pp. 271--278 (02 2017).
  \doi{http://doi.org/10.14794/ICAI.10.2017.63},
  \url{http://icai.uni-eszterhazy.hu/icai2017/uploads/papers/2017/final/ICAI.10.2017.63.pdf}

\bibitem{Carruth2016High}
Carruth, C.: High performance code 201: Hybrid data structures (09 2016),
  \url{http://youtube.com/watch?v=vElZc6zSIXM}, (accessed 2020.10.26.)

\bibitem{chess2004static}
Chess, B., McGraw, G.: Static analysis for security. IEEE security \& privacy
  \textbf{2}(6),  76--79 (2004)

\bibitem{Chrome}
Cimpanu, C.: Chrome: 70\% of all security bugs are memory safety issues (2020),
  \url{http://www.zdnet.com/article/chrome-70-of-all-security-bugs-are-memory-safety-issues/},
  (accessed 2020.12.21.)

\bibitem{CallEvent}
{Clang Doxygen}: \texttt{clang$::$ento$::$CallEvent} Class Reference (2020),
  \url{http://clang.llvm.org/doxygen/classclang_1_1ento_1_1CallEvent.html},
  (accessed 2020.12.21.)

\bibitem{CheckerDoc}
{Clang Doxygen}: \texttt{clang$::$ento$::$CheckerDocumentation} Class Reference
  (2020),
  \url{http://clang.llvm.org/doxygen/classclang_1_1ento_1_1CheckerDocumentation.html},
  (accessed 2020.12.21.)

\bibitem{SVal}
{Clang Doxygen}: \texttt{clang$::$ento$::$SVal} Class Reference (2020),
  \url{http://clang.llvm.org/doxygen/classclang_1_1ento_1_1SVal.html},
  (accessed 2020.12.21.)

\bibitem{dergachev2016clang}
Dergachev, A.: Clang static analyzer: A checker developer’s guide. (2016)

\bibitem{CodeChecker}
{Ericsson AB.}: {C}ode{C}hecker, \url{http://github.com/Ericsson/CodeChecker},
  (accessed 2020.10.26.)

\bibitem{ChromeAnalyze}
{Google Inc.}: The Clang Static Analyzer (2020),
  \url{http://chromium.googlesource.com/chromium/src/+/master/docs/clang_static_analyzer.md},
  (accessed 2020.12.21.)

\bibitem{grossman2005cyclone}
Grossman, D., Hicks, M., Jim, T., Morrisett, G.: Cyclone: A type-safe dialect
  of c. C/C++ Users Journal  \textbf{23}(1),  112--139 (2005)

\bibitem{Hampapuram2005}
Hampapuram, H., Yang, Y., Das, M.: Symbolic path simulation in path-sensitive
  dataflow analysis. SIGSOFT Softw. Eng. Notes  \textbf{31}(1),  52--58 (Sep
  2005). \doi{10.1145/1108768.1108808},
  \url{http://doi.acm.org/10.1145/1108768.1108808}

\bibitem{Horvath2014Szemantikus}
Horv\'{a}th, G.: Szemantikus keresés {\CC{}} kódbázisban. Bachelor's thesis,
  E\"{o}tv\"{o}s Lor\'{a}nd University, Faculty of Informatics (2014), (title
  in English: \emph{Semantic search in {\CC{}} code-bases})

\bibitem{StdDraft}
ISO/IEC: Working Draft, Standard for Programming Language {C}++, N4868 (2020),
  \url{http://www.open-std.org/jtc1/sc22/wg21/docs/papers/2020/n4868.pdf},
  (accessed 2020.12.21.)

\bibitem{johnson2013don}
Johnson, B., Song, Y., Murphy-Hill, E., Bowdidge, R.: Why don't software
  developers use static analysis tools to find bugs? In: 2013 35th
  International Conference on Software Engineering (ICSE). pp. 672--681. IEEE
  (2013)

\bibitem{king1976symbolic}
King, J.C.: Symbolic execution and program testing. Communications of the ACM
  \textbf{19}(7),  385--394 (1976)

\bibitem{kovacs2019detecting}
Kov{\'a}cs, R., Horv{\'a}th, G., Porkol{\'a}b, Z.: Detecting {\CC{}} lifetime
  errors with symbolic execution. In: Proceedings of the 9th Balkan Conference
  on Informatics. pp.~1--6 (2019)

\bibitem{kremenek2008finding}
Kremenek, T.: Finding software bugs with the {C}lang {S}tatic {A}nalyzer. Apple
  Inc  (2008)

\bibitem{lattner2008llvm}
Lattner, C.: {LLVM} and {C}lang: Next generation compiler technology. In: The
  BSD conference. vol.~5 (2008)

\bibitem{lattner2004llvm}
Lattner, C., Adve, V.: {LLVM}: A compilation framework for lifelong program
  analysis \& transformation. In: International Symposium on Code Generation
  and Optimization, 2004. CGO 2004. pp. 75--86. IEEE (2004)

\bibitem{ASTMatchers}
{LLVM/Clang}: {AST} {M}atcher {R}eference,
  \url{http://clang.llvm.org/docs/LibASTMatchersReference.html}, (accessed
  2020.10.26.)

\bibitem{madsen2015static}
Madsen, M.: Static analysis of dynamic languages.
  http://pure.au.dk/ws/files/85299449/Thesis.pdf  (2015)

\bibitem{CppCheck}
Marjam{\"a}ki, D.: {CppCheck} - a tool for static {C/{\CC{}}} analysis (2020),
  \url{http://cppcheck.sourceforge.net/}, (accessed 2020.12.21.)

\bibitem{Zhivich2009}
Michael, Z., Robert, K.C.: The real cost of software errors. IEEE Security \&
  Privacy  \textbf{7}(2),  87--90 (2009). \doi{10.1109/MSP.2009.56},
  \url{http://dx.doi.org/10.1109/MSP.2009.56}

\bibitem{moller2012static}
M{\o}ller, A., Schwartzbach, M.I.: Static program analysis. Notes. Feb  (2012)

\bibitem{10.1145/1273442.1250746}
Nethercote, N., Seward, J.: Valgrind: A framework for heavyweight dynamic
  binary instrumentation. SIGPLAN Not.  \textbf{42}(6),  89–100 (Jun 2007).
  \doi{10.1145/1273442.1250746}, \url{http://doi.org/10.1145/1273442.1250746}

\bibitem{Rice:53}
Rice, H.G.: Classes of recursively enumerable sets and their decision problems.
  Trans. Amer. Math. Soc.  \textbf{74},  358--366 (1953)

\bibitem{Szalay2020Towards}
Szalay, R.: Towards decoupling nullability semantics from indirect access in
  pointer use. In: Kov\'{a}sznai, G., Fazekas, I., T\'{o}m\'{a}cs, T. (eds.)
  {P}roceedings of the 11th {I}nternational {C}onference on {A}pplied
  {I}nformatics. pp. 328--340. Eszterh{\'{a}}zy K{\'{a}}roly University, Eger,
  Hungary (01 2020), \url{http://ceur-ws.org/Vol-2650/paper34.pdf}

\bibitem{static-analyzer}
{The LLVM Foundation}: {Clang Static Analyzer} (2019),
  \url{http://clang-analyzer.llvm.org}, (accessed 2019.02.28.)

\bibitem{clangtidy}
{The LLVM Foundation}: {Clang-Tidy} (2019),
  \url{http://clang.llvm.org/extra/clang-tidy}, (accessed 2019.02.28.)

\bibitem{ClangDoc}
{The LLVM Foundation}: Getting Started: Building and Running Clang (2020),
  \url{http://clang.llvm.org/get_started.html}, (accessed 2020.12.21.)

\bibitem{Microsoft}
Thomas, G.: A proactive approach to more secure code (2019),
  \url{http://msrc-blog.microsoft.com/2019/07/16/a-proactive-approach-to-more-secure-code/},
  (accessed 2020.12.21.)

\bibitem{turing1936computable}
Turing, A.M., et~al.: On computable numbers, with an application to the
  entscheidungsproblem. J. of Math  \textbf{58}(345-363), ~5 (1936)

\bibitem{CppQuery}
{Xazax-hun}: {CppQuery} (2014), \url{http://github.com/Xazax-hun/CppQuery},
  (accessed 2020.10.26.)

\bibitem{xu2010memory}
Xu, Z., Kremenek, T., Zhang, J.: A memory model for static analysis of c
  programs. In: International Symposium On Leveraging Applications of Formal
  Methods, Verification and Validation. pp. 535--548. Springer (2010)

\end{thebibliography}
\end{document}